\documentclass[]{aa}
\usepackage{amsmath,latexsym,amssymb,amsbsy}
\usepackage{graphicx}
\usepackage{float}
\usepackage{epstopdf}
\usepackage{multirow}
\usepackage{natbib}
\usepackage{multicol}
 
\usepackage{soul, xcolor}

\usepackage[citecolor=blue,
            anchorcolor = blue,
            colorlinks=true,
            linkcolor = blue,
            urlcolor=blue,
            breaklinks=true]{hyperref} 

\begin{document}

\title{{\bf Oscillations of cometary tails: a vortex shedding phenomenon?}}
\author{G. Nistic\`o\inst{1,2} \and V. Vladimirov\inst{1} \and V. M. Nakariakov\inst{1} \and K. Battams\inst{3} \and V. Bothmer\inst{2}}

\institute{Centre for Fusion, Space and Astrophysics, University of Warwick, Coventry, UK \and Institut f\"ur Astrophysik, Georg-August-Universit\"at, G\"ottingen, Germany \and Naval Research Laboratory, Washington (DC), USA}

\authorrunning{Nistic\`o et al.}
\titlerunning{Oscillations in cometary tails}


\abstract
{During their journey to perihelion, comets may appear in the field-of-view of space-borne optical instruments, showing in some cases a nicely developed plasma tail extending from their coma and exhibiting an oscillatory behaviour.}
{The oscillations of cometary tails may be explained in terms of vortex shedding because of the interaction of the comet with the solar wind streams. Therefore, it is possible to exploit these oscillations in order to infer the value of the Strouhal number $St$, which quantifies the vortex shedding phenomenon, and the physical properties of the local medium.}
{We used the Heliospheric Imager (HI) data of the Solar TErrestrial Relations Observatory (STEREO) mission to study the oscillations of the tails of the comets  2P/Encke and C/2012 S1 (ISON) during their perihelion in Nov 2013, determining the Strouhal numbers from the estimates of the halo size, the relative speed of the solar wind flow and the period of the oscillations.}
{We found that the estimated Strouhal numbers are very small, and the typical value of $St\sim0.2$ would be extrapolated for size of the halo larger than $\sim10^6$ km.}
{Despite the vortex shedding phenomenon has not been unambiguously revealed, the findings suggest that some MHD instability process is responsible for the observed behaviour of cometary tails, which can be exploited for probing the physical conditions of the near-Sun region.}
\maketitle

\section{Introduction}
Optical instruments aboard space missions, like the Solar Heliospheric Observatory (SoHO)/LASCO \citep{Brueckner1995} 
and the Solar TErrestrial RElations Observatory (STEREO)/SECCHI coronagraphs \citep{Kaiser2005} have returned 
observations of more than 3,200 new and previously known comets \citep{Battams2016}. More than 85\% of these 
have a perihelion very close to the Sun, and are defined as \lq\lq sungrazing\rq\rq~comets. Usually, they disappear 
before reaching their perihelion, as a result of fragmentation and vaporisation at distances of typically 6--10 solar radii \citep{Knight2012, Biesecker2002}. 
However, few exceptional cases of comets flying inside the solar corona and observed by extreme ultra-violet 
(EUV) imagers \citep{Schrijver2012, Downs2013,McCauley2013} have been reported. 
Therefore, one area of interest in comets is related to the possibility of exploiting them as natural probes for the solar corona and near-Sun environment \citep{Ramanjooloo}. 
A tail of ions from the cometary nuclei is formed, which 
interacts with the local medium exhibiting a swaying-like motion, as also evident with the first observation of the comet 2P/Encke in 2007 with the Heliospheric Imager (HI) 1 of STEREO-A \citep{Vourlidas2007}. 
The features observed in the Encke's tail have been interpreted in terms of turbulent eddies rooted in the solar wind and traced by the cometary plasma \citep{DeForest2015}. On the other hand, the observed comet-solar wind system is more analogous to that of an object of finite size immersed in a flow with a K\'arm\'an vortex street formed in the wake of the obstacle. 
The phenomenon of vortex shedding has been widely invoked both in science and engineering. 
In solar physics it has been used to explain the excitation and selectivity of kink oscillations in coronal loops \citep{Nakariakov2009} and global oscillations of halo CMEs measured with coronagraphs \citep{Lee2015}. 
Morover, vortices due to Kelvin-Helmholtz instability have been observed at the flanks of an expanding CME
 \citep{Foullon2011}. In the context of comets, the interaction of the solar wind with the cometary halo 
may lead to the formation of shed vortices: the ion tail would periodically oscillates as a consequence of the appearance of a periodic force caused by the succession of eddies with opposite vorticity, similarly to flags waving in a wind.
The fluid behaviour past an obstacle is described by the well-known Reynolds number $Re = VL/\nu$ 
(with $V$ the relative flow speed, $L$ the obstacle size, and $\nu$ the kinematic viscosity), 
and the Strouhal number $St = L/(P V)$, which takes into account the period $P$ of the shed vortices. 
The relationship between them is not unambiguously established \citep{Sakamoto1990,Ponta2004}, 
but it can be used for the estimation of the kinematic viscosity of the fluid. 
Here, we aim to analyse the dynamics of the tails of 2P/Encke and the sungrazing comet C/2012 S1 (ISON) 
observed with the HI-1 and 2 of STEREO-A during their perihelion in 2013 in the context of the vortex shedding phenomenon. The paper is organised as follows: Section \ref{sec:obs} presents the overall observations; values of the $St$ numbers (and the associated $Re$ ones) from the estimates of the halo size, 
the relative speed of the solar wind flow, and the properties of the observed oscillations 
(wavelength, period, amplitude) of the tails are shown in Sect. \ref{sec:ana}; discussion and conclusions in Sect. \ref{sec:disc}. We demonstrate how these observations can be exploited to determine 
the physical properties of the solar wind plasma.

\section{Observations and data}
 \label{sec:obs}
 
\begin{figure*}[htpb]
\centering
	\begin{tabular}{c c c}
		\includegraphics[width=6 cm]{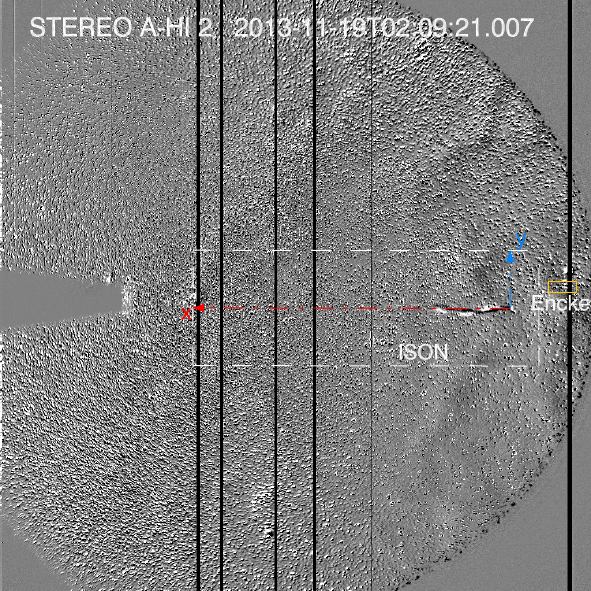} &
		\includegraphics[width=6 cm]{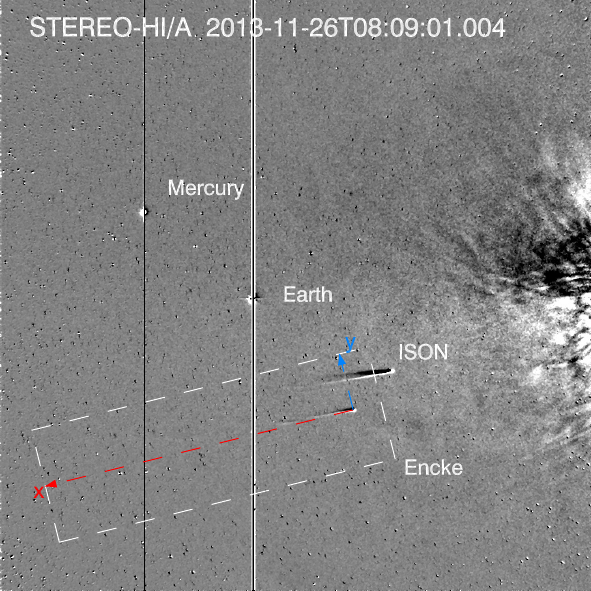} &
		\includegraphics[width=6 cm]{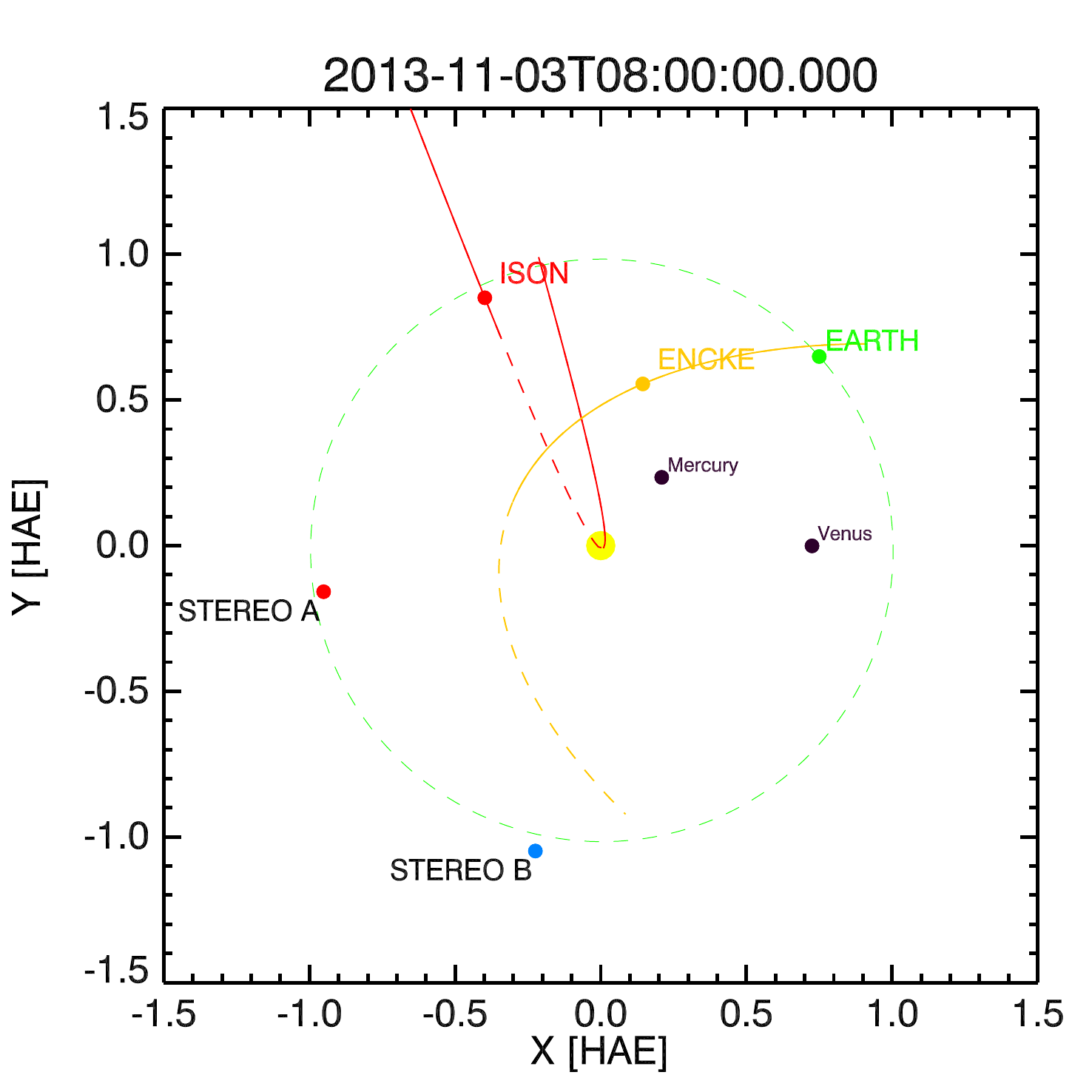} \\
	\end{tabular}
	\caption{Running difference of Encke and ISON from STEREO-A/HI-2 (left) and HI-1 (middle). Orbits of the comets Encke (yellow line) and ISON (red line) in the Heliocentric-Aries-Ecliptic (HAE) coordinate system (right). Dashed lines indicate where the trajectory is below the ecliptic plane.  The Earth is the green dot (with the associated orbit dashed line style), while STEREO-A and B are the red and blue dots, respectively. For reference, the position of the inner planets is also shown. The corresponding animations for each panel are available online.}
	\label{fig1}
\end{figure*}

The HI-1 telescope of STEREO A provides white-light images of the inner heliosphere 
covering a field-of-view between $\sim$4$^{\circ}$ and $\sim$24 solar elongation angle 
from the East solar limb ($\sim 15-84~R_\odot$), with a pixel size of 1.2 arcmin ($\sim$72 arcsec) 
\citep{Horward2008}, while HI-2 observes the outer heliosphere with an 
angular range of $\sim19^\circ-\sim89^\circ$ ($66-318~R_\odot$), with an image pixel size of 4.3 arcmin.
  The typical cadence of each instrument is 40 and 120 min, respectively, with expsoure time of typically 40-minute. 
We have used Level 2 FITS files from the UK Solar System Data Centre
\footnote{\url{http://www.ukssdc.ac.uk/solar/stereo/data.html}.}, based on 1-day background subtracted for HI-1, 
and 3-day background subtracted for HI-2 to remove the excess of the F-corona brightness and stray light, and covering a time interval between 05 Nov and 9 Dec 2013. Then, we read the FITS files using the routine {\tt mreadfits}, which is part of SolarSoftWare (SSW) \footnote{Set of integrated software libraries and system utilities based on the Interactive Data Language (IDL): \url{http://www.lmsal.com/solarsoft/}.}, and obtained the corresponding  headers and image arrays with size of 1024$\times$1024 pixels.

\begin{figure*}
\centering
			\includegraphics[width=16 cm]{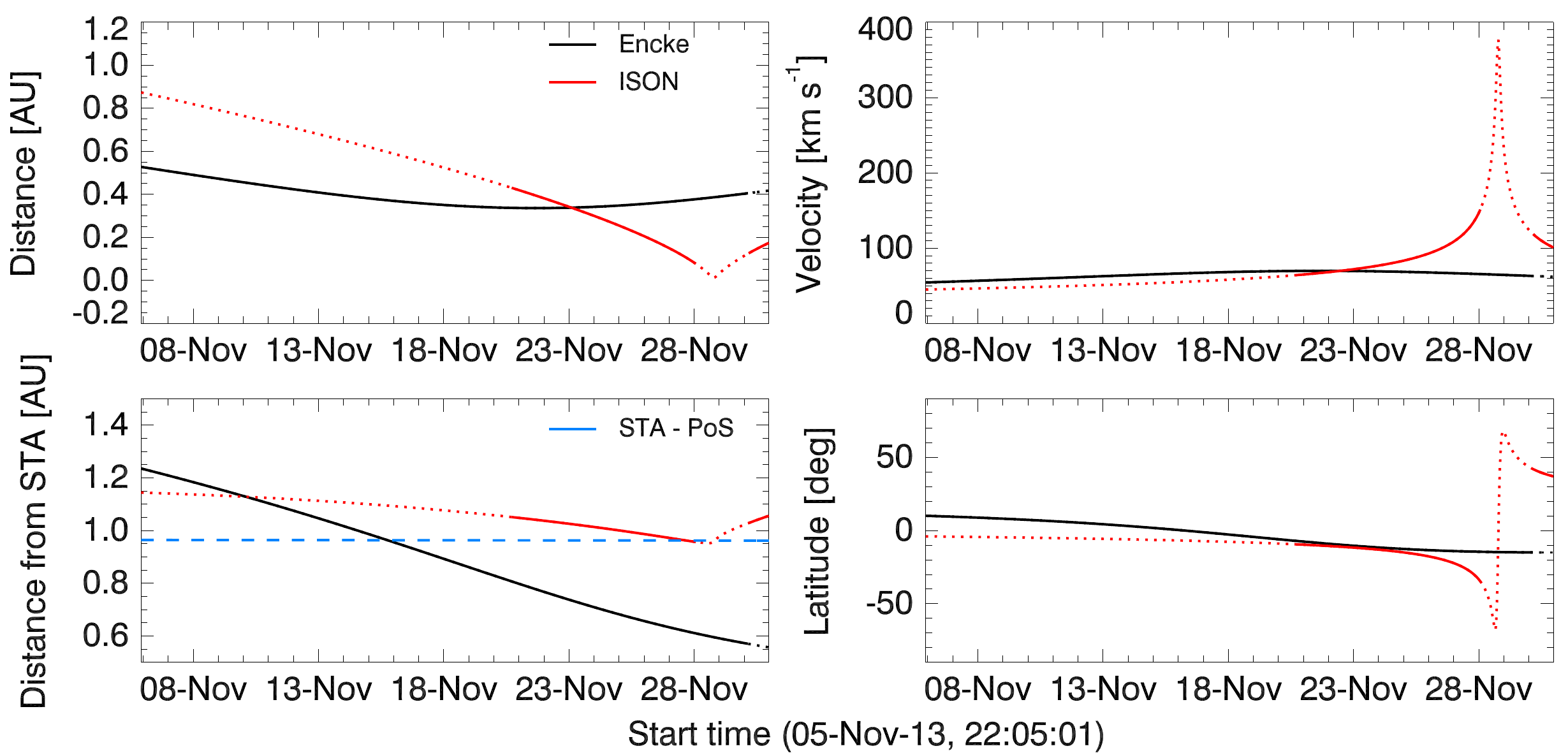}
\caption{Left: variation of the distance from the Sun (top) for Encke (black) and ISON (red), and from STEREO-A in AU (bottom). Continuous lines mark the time intervals when the comets are visible in the HI-1 FoV, dotted lines the opposite. The blue dashed line in the bottom-panel indicates the STEREO A-Sun distance, which is a reference line to visualise when the comets are behind or above the plane-of-sky.)  Right: Orbital speed of the comets (top), and variation of the latitude in the Helio-Centric-Inertial (HCI) coordinate system (bottom).}
				\label{fig2}
	\end{figure*}   

To better reveal the fluctuations of the tails, HI images have been processed for background stars removal by cross-correlating each pair of consecutive images from our dataset: we found the relative pixel shift between them, translated the subtrahend image of this amount, and finally performed the difference. An example of images is given in Fig. \ref{fig1}. 
    The positions of Encke and ISON during the entire time of the observations are found by de-projecting the ephemerides of the comets to the STEREO-A/HI images using the routines {\tt fitshead2wcs} and {\tt wcs\_get\_pixel} of the World Coordinate Systems (WCS) package, which is included in SSW \citep[][]{Thompson2010}. The ephemerides are initially read and processed within SPICE, which is part of the Navigation and Ancillary Information Facility (NAIF) and also implemented in SSW with the SUNSPICE package. SPICE kernels of the comets (i.e. files in {\tt .bsp} format storing the ephemerides)  are downloaded from the following website \url{http://ssd.jpl.nasa.gov/x/spk.html},
and load by the {\tt cspice\_furnsh} routine. Position and velocity in a desired coordinate system are obtained with {\tt get\_sunspice\_coord}, and then used to make plots in Fig. \ref{fig1}-right panel, and Fig. \ref{fig2}. 
Then, we created a series of running difference sub-images with a new reference frame co-moving with each 
single comet (Fig. \ref{fig1}): the comet's head is fixed, while the tail almost lies along the horizontal axis. 
The orbital properties of Encke and ISON are different (Fig. \ref{fig1}-right and \ref{fig2}): 
Encke reached the perihelion at 0.33 AU on the 21th Nov 2013 with an orbital speed of $\sim$70 km/s, 
and at the time of the observations its orbit was pretty close to the solar equatorial plane 
(between approximately -10$^\circ$ and +10$^\circ$ in latitude). 
On the contrary ISON orbited along a hyperbolic trajectory, spanning several degrees 
in latitude at the perihelion with the closest distance at 0.01 AU from the Sun's centre 
(just only $\sim1.15 R_\odot$ from the solar photosphere) reached on the 28th Nov 2013, 
with an orbital velocity of almost 400 km s$^{-1}$. 
However, when observed with HI-1 (approximately until the 26th November), 
both comets have similar orbital speeds and latitudes, moving out of the plane 
of observations of the instrument ($\sim0.95 AU$, dashed-blue line in the bottom-left panel of Fig. \ref{fig2}): 
the distance of ISON from STEREO A ranges between $\sim 1.15-0.95$ AU, 
while that of Encke between $\sim1.2-\sim0.6$ AU during the time of the observations.

\begin{figure*}[htpb]
\centering
	\begin{tabular}{c c}
		\includegraphics[width=7 cm]{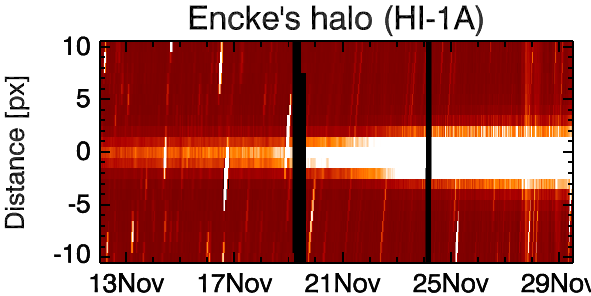} &
		\includegraphics[width=7 cm]{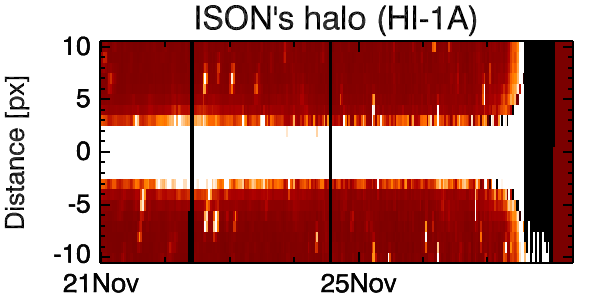} \\
		\includegraphics[width=7 cm]{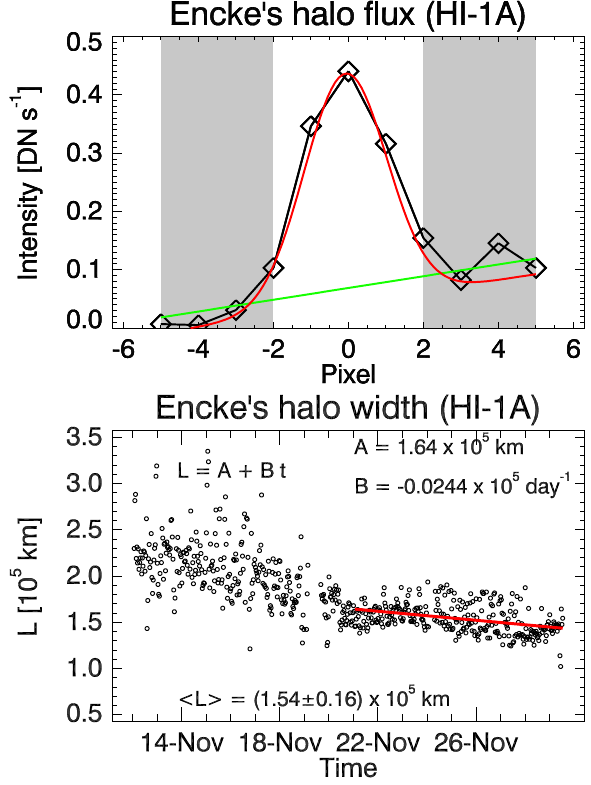} &
		\includegraphics[width=7 cm]{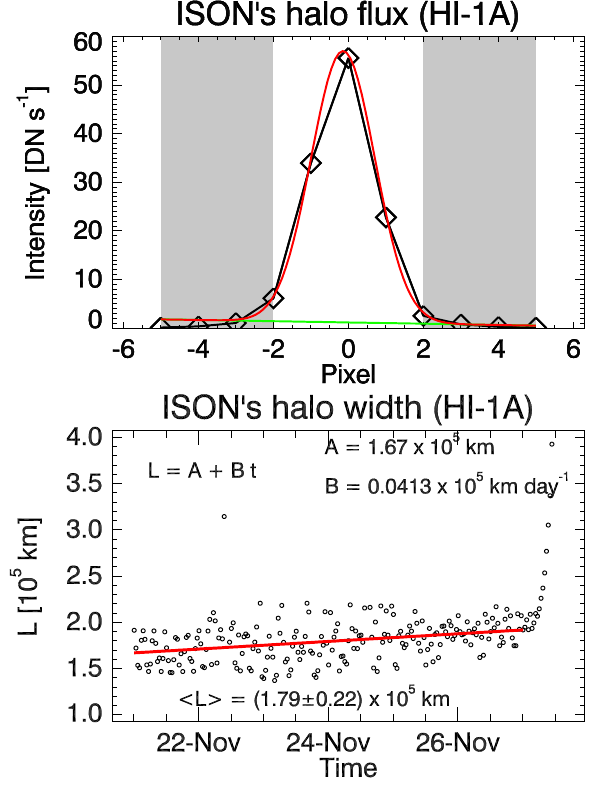} \\
	\end{tabular}
	\caption{Time-distance maps (top), example of intensity profile with related fittings (middle), and $L$ vs. time for the Encke (left) and ISON's (right) halo with HI-1 A.}
	\label{fig_coma}
\end{figure*}

\section{Analysis} 
\label{sec:ana}

	 To quantity the Strouhal number, we have to estimate typical values of the size of the cometary halo $L$, which we assume to play the role of an obstacle immersed in the solar wind flow, the relative speed comet-solar wind $V$, and the period $P$ of the tail oscillation which we assume equivalent to that of the hypothetical shed vortices.
	
	\subsection{Determination of the halo size}
	The size of the halos is inferred by constructing time distance maps from the normal intensity images
	 with a vertical slit across the comet's head (Fig. \ref{fig_coma}) in the processed HI-1 sub-images.
	 The horizontal bright feature at the centre of the time distance maps is the signature of the comet's 
	 halo. We have determined the size by fitting the intensity profile with a Gaussian function at each time 
	 \citep[we used the MPFIT routines by ][]{Markwardt2009}.
	 The intensity profile across the halo is sampled over 11 pixels across: 4 pixels at the sides of this
	 spatial interval are taken as a background (shaded region in the middle panels of Fig. \ref{fig_coma}),
	 which is fitted with a linear function added to the Gaussian function (red). The full-width of the
	 Gaussian function at the background intensity level is chosen as a good approximation for the apparent
	 size of the halo $L'_\mathrm{pix}(t) = 2\sqrt{2\ln(I_\mathrm{max}(t)/I_\mathrm{back})}\sigma(t)$, where $I_\mathrm{max}(t)$ is
	 the height of the peak intensity of the coma, and $I_\mathrm{back}$ the average value of the background
	 intensity of $\sim$0.1 DN s$^{-1}$. Measurements are strongly affected by the point-spread function
	 (PSF) of HI-A, which is estimated of the order of $w_\mathrm{PSF}=1.48-1.69$ pixel \citep{Bewsher2010}.
	 In addition, the limiting magnitude for HI-1 is approximately 13.5.
	  In a way similar to what shown by \citet{Aschwanden2008} for coronal loops, the effective size of the coma measured in pixel units $L'_\mathrm{pix}$ is given by
		\begin{equation}
			L'_\mathrm{pix} = \sqrt{L_\mathrm{pix}^2 + w_\mathrm{PSF}^2}, 
			\label{L_eq}
		\end{equation}
		which is used to determine $L_\mathrm{pix}$.
		These values are converted into physical units by considering the the radius of the Sun in arcsec (retrieved from the header under the keyword {\tt RSUN}), and the CCD plate scale 
		$\Delta_\mathrm{pix} \approx 0.02$ deg pix$^{-1} \approx 0.5 \times 10^5$ km pix$^{-1}$({\tt CDELT1} keyword), both defined at the STEREO A-Sun
		distance
		($d_\mathrm{Sun}$, obtained from the header keyword {\tt DSUN\_OBS} ) and corrected for the relative distance $d_\mathrm{C}$ comet-observer (calculated via {\tt get\_sunspice\_lonlat}).
		Therefore, the size of the halo is found as $L = L_\mathrm{pix}~\Delta_\mathrm{pix}~d_\mathrm{C}/d_\mathrm{Sun}$. 
	The data points for both comets are fitted with a linear function (red line in the bottom panels of Fig. \ref{fig_coma}): ISON presents a clear increase of the halo size over time, while the Encke's halo size is slightly decreasing (we have not considered the broader cloud of data points since these values are affected by the low contrast between the Encke's brightness and the background). Average values of the halo size are found to be $L_\mathrm{Encke} = (1.54 \pm 0.16)\times 10^5$ km, 
	and $L_\mathrm{ISON} = (1.79\pm 0.22)\times10^5$ km. These estimates are consistent with typical values of 
	cometary halos/comas found in literature, which can also reach values of $10^6-10^7$ km \citep{Ramanjooloo}.

	\subsection{Determination of the relative speed}
\begin{figure*}[htpb]
	\centering
	\begin{tabular}{c c}
	\includegraphics[width = 9 cm]{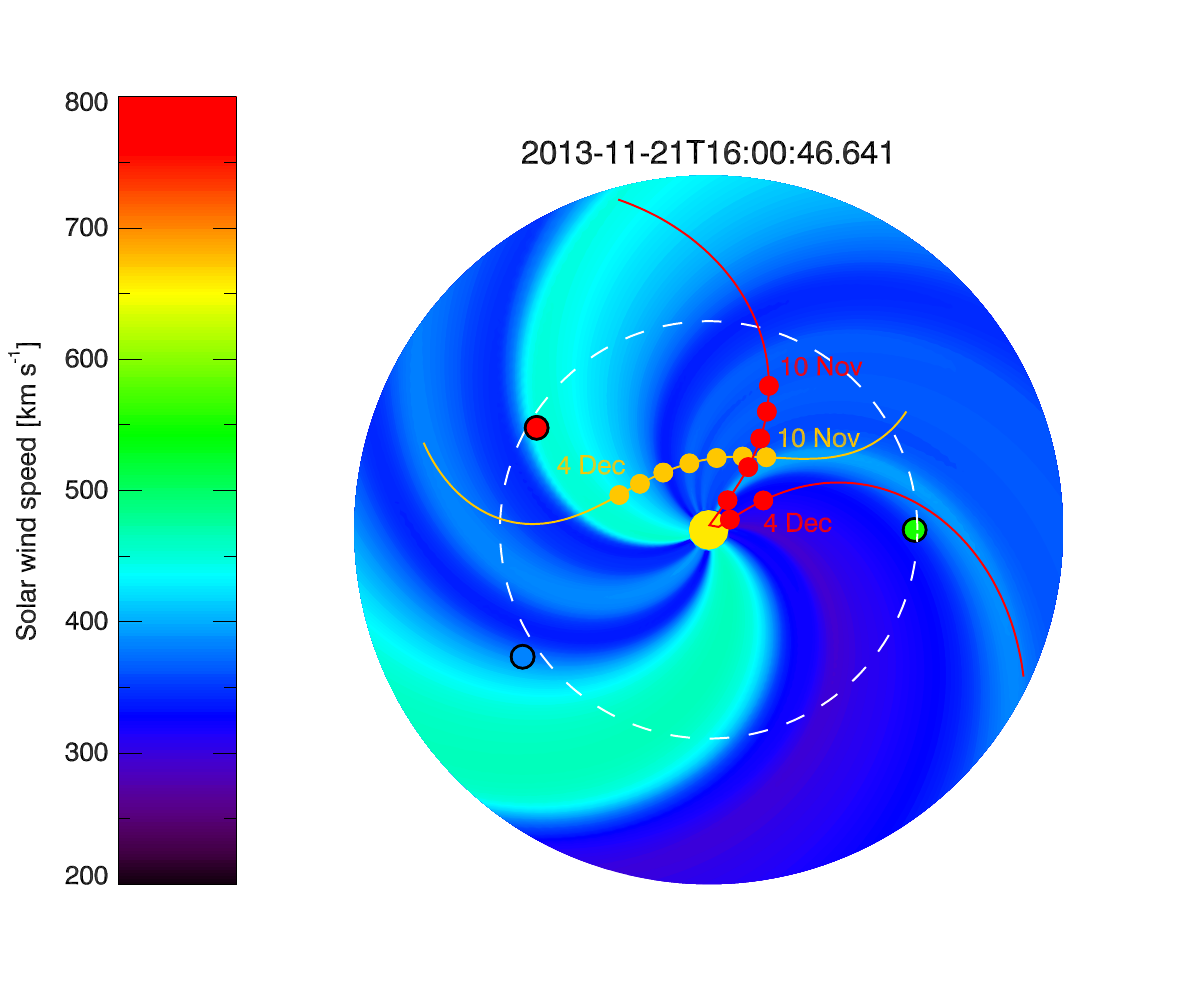} &
	\includegraphics[width = 7 cm]{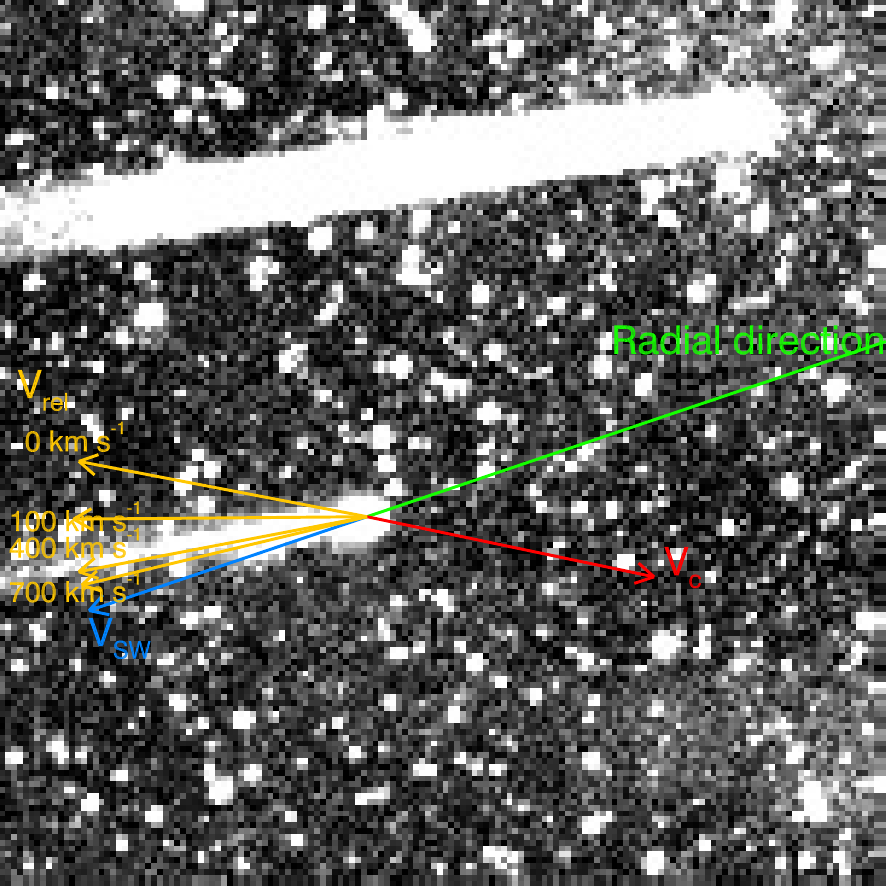} \\
	 \includegraphics[width = 9 cm]{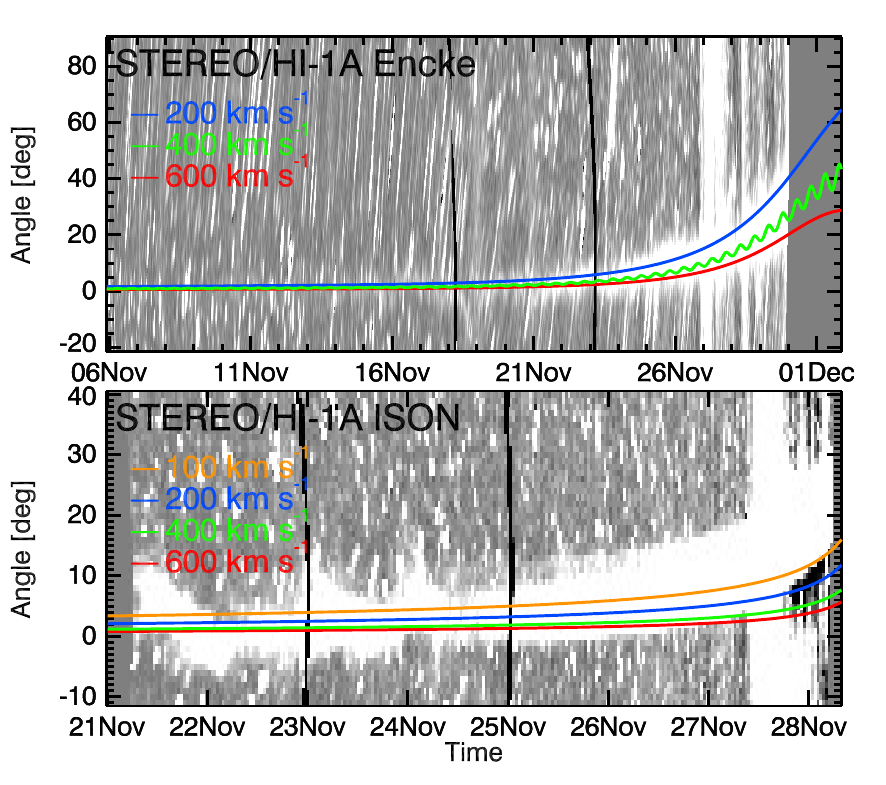} &
	\includegraphics[width = 9 cm]{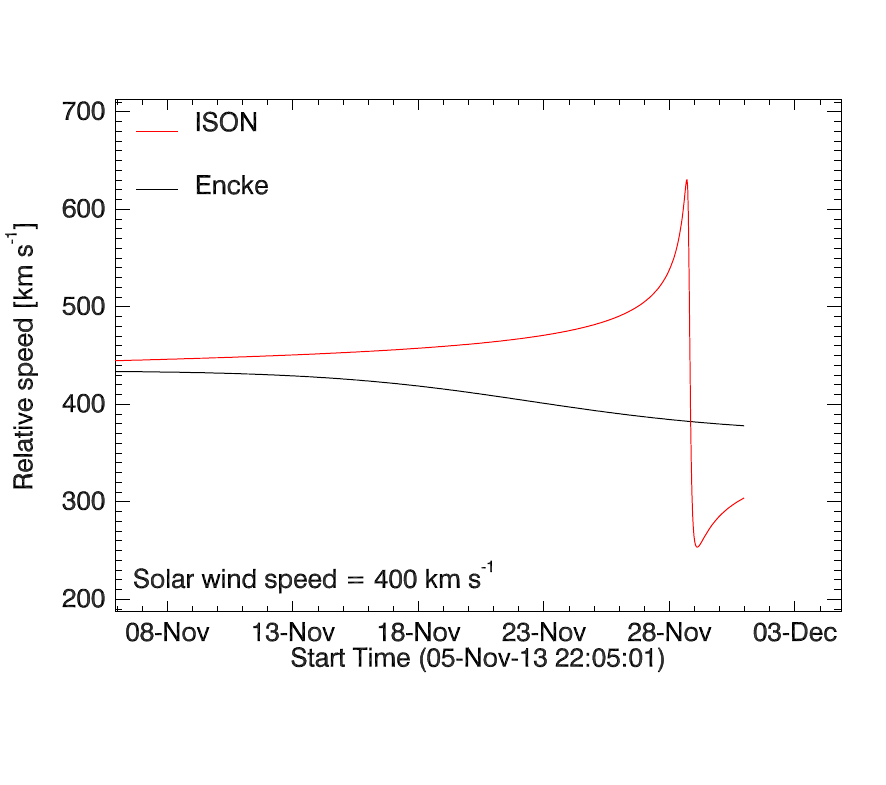} 	\\

	\end{tabular}
	\caption{Top right: Solar wind speed on the solar equatorial plane provided by the ENLIL model, 
	with the position of the Earth, STEREO A and B, and the trajectories of the comets Encke (orange) 
	and ISON (red). The corresponding movie is available online. Top left: image showing the projected radial direction from the Sun, the projected direction of the
	orbital speed for Encke (red), and the resulting vectors $V$ (yellow) 
	determined as a vectorial sum between the solar wind $V_\mathrm{SW}$ and comet speed $V_\mathrm{C}$ vectors. 
	For a solar wind speed of 400 km s$^{-1}$, the projected direction of $V$ almost coincides with 
	the tail direction. Bottom left: Time distance maps for Encke (right) and ISON (left) showing the 
	inclination of the tail (the bright feature) with respect to the projected radial direction, 
	and compared for different profiles of the aberration angle. In one case, we show 
	the behaviour of the aberration angle profile for Encke assuming a sinusoidally variable solar wind with mean value 
	of 400 km s$^{-1}$, amplitude of 50 km s${^{-1}}$ and period of 12 hr. 
	Bottom right: plot of the relative speed magnitude $V$ vs. time of observations assuming a constant and radial solar wind speed of 400 km s$^{-1}$.}
	\label{fig_speed}	
	\end{figure*}

	Values of the relative speed of the solar wind past the halos strongly
	depend on the comet orbits and the intrinsic variable nature of the solar wind speed, 
	which ranges between 300 (slow wind) and 800 km s$^{-1}$ (fast wind). 
	Accurate knowledge of the solar wind speed at the positions of the comets would require 
	forward modelling or extrapolations based upon the conditions of the solar corona 
	and/or satellite measurements. Figure \ref{fig_speed}-top-left shows 
	the solar wind speed on the solar equatorial plane provided by the ENLIL model \citep{Odstrcil2003} at 
	the time of the Encke's perihelion in the Helio-Earth-EQuatorial (HEEQ) coordinate system. 
	The Earth position is fixed and represented with a green dot, while STEREO A and B 
	are given with red and blue dots, respectively. The trajectories of Encke and ISON projected on this 
	plane are shown in orange and red, respectively, 
	with some dots showing the positions of the comets with an 
	interval of 4 days between the 10th Nov and the 4th Dec. Both comets seems to cross different 
	solar wind streams that have speeds between 300--450 km s$^{-1}$. 
	The relative speed $V$ is determined by the vectorial sum of the solar wind flow $V_\mathrm{SW}$ 
	and the orbital comet speed $V_\mathrm{C}$, that is ${\bf V} = {\bf V}_\mathrm{SW} - {\bf V}_\mathrm{C}$, and the plasma tail 
	should extend along this resulting vector, which forms an angle with the solar wind direction defined as
	the aberration angle (Fig. \ref{fig_speed}-top-right).
	The aberration angle $\alpha$ between the relative speed vector
	and the radial direction is defined as $\alpha = \cos^{-1} \left(\frac{{\bf V} \cdot {\bf V}_\mathrm{SW}}{|{\bf
	V}| |{\bf V}_\mathrm{SW}|}\right)$. Therefore, given ${\bf V}_\mathrm{C}$, we determine the function $\alpha = \alpha(t, V_\mathrm{{SW}})$ in
	the time range of the observations and for different values of $V_\mathrm{SW}$ (200, 400, 600, ... km s$^{-1}$),
	and visually compare the hypothetical aberration angle profiles (de-projected according to the STEREO A view) with the location of the tail inferred
	from a TD map. The TD maps in Fig. \ref{fig_speed} show the normal intensity extracted from a semi-circular slit
	located at 20 and 60 pixels from the coma centre of Encke and ISON, respectively. The 0 in the vertical
	axis coincides with the projected radial direction comet-Sun. 
	The aberration angle profiles are overplotted for different values of the radial solar wind speed $V_{SW}$. When both comets are relatively far from their
	perihelion,
	the different aberration profiles tend to coincide because of projection effects (the tails extend along
	the apparent radial direction). Close to perihelion, the tails undergo a considerable angular deviation,
	which is well-fitted by a solar wind speed of 400 km s$^{-1}$ in the case of Encke. The same is not
	unambiguously clear for ISON, and the position of the tail may be affected by other factors, like the
	hyperbolic orbit of the comet, 
	spurious projection effects, the nature of the solar wind out of the equatorial plane, or the composition of
	the ISON's tail (e.g. strong percentage of dust particles) which would affect the direction. Despite this, we tend to consider an average solar wind flow $V_\mathrm{SW}=400$ km s$^{-1}$ even for ISON.
	It is interesting to notice that periodic changes in the solar wind speed can determine periodic variation of the aberration angle, hence oscillations of the cometary tails (see the green line in Fig. \ref{fig_speed}-middle left). The green oscillatory pattern in the Encke's TD map is obtained with an amplitude velocity of 50 km s$^{-1}$ (of the order of the Alfv\'en speed in the solar wind). However, the oscillations are not evident when the projected aberration and radial direction coincide.
	In addition, other parameters like solar wind density or the magnetic field vector could somehow influence the observed oscillations, but we do not consider any quantified contribution in the present study.
	
	After having defined the more probable
	solar wind speed (in our case 400 km s$^{-1}$), the magnitude of the relative flow is found as $V =
	\sqrt{V_\mathrm{SW}^2 + V_\mathrm{C}^2 - 2 ({\bf V}_\mathrm{SW} \cdot {\bf V}_\mathrm{C})}$.
	Figure \ref{fig_speed}-bottom-right shows the profile of the relative speed $V$ for Encke and ISON 
	during the observations: the relative speed for Encke is approximately limited 
	between 380--440 km s$^{-1}$, while ISON reaches values up to $\sim$ 650 km s$^{-1}$, at the perihelion.

	\subsection{Determination of the period}
\begin{figure*}[htpb]
\centering
	\begin{tabular}{c c}
		\includegraphics[width = 0.5\textwidth]{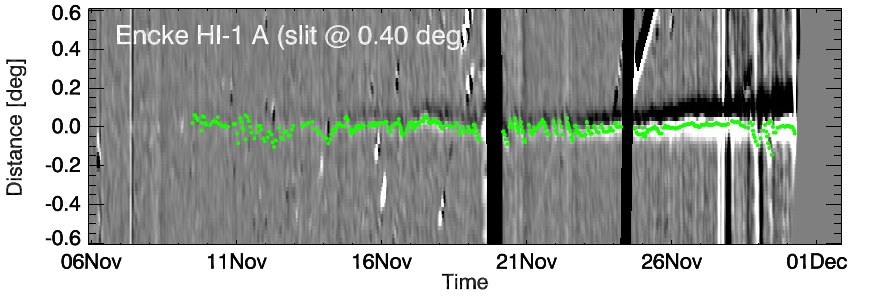} &
		\includegraphics[width = 0.5\textwidth]{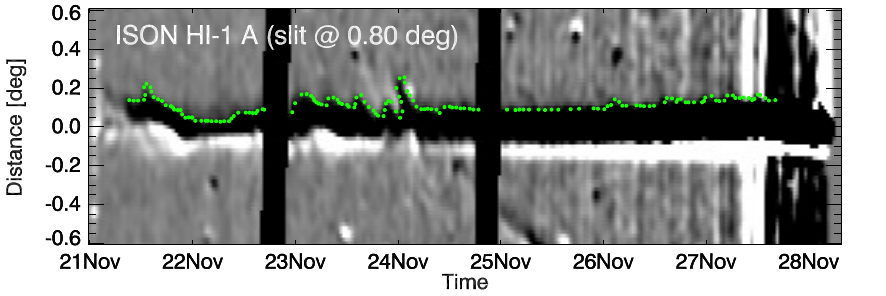} \\
		\includegraphics[width = 0.5\textwidth]{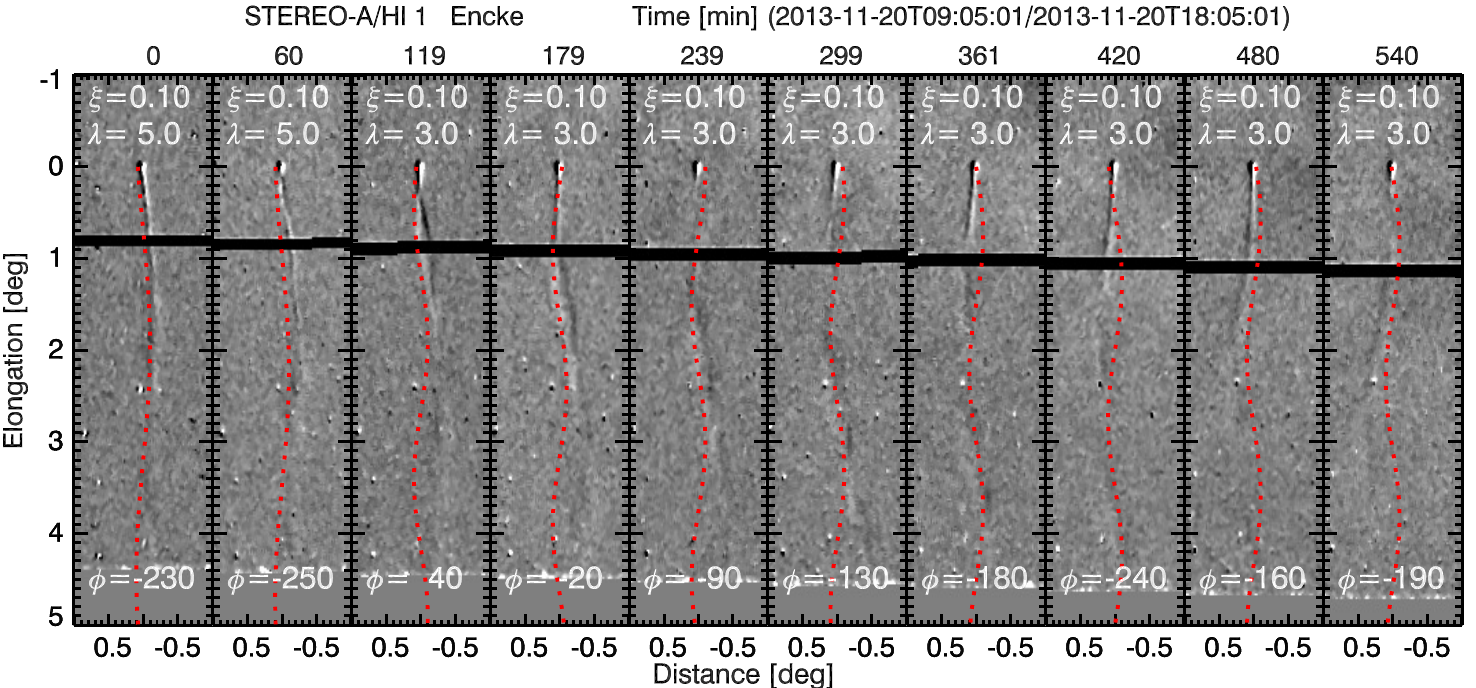} &
		\includegraphics[width = 0.5\textwidth]{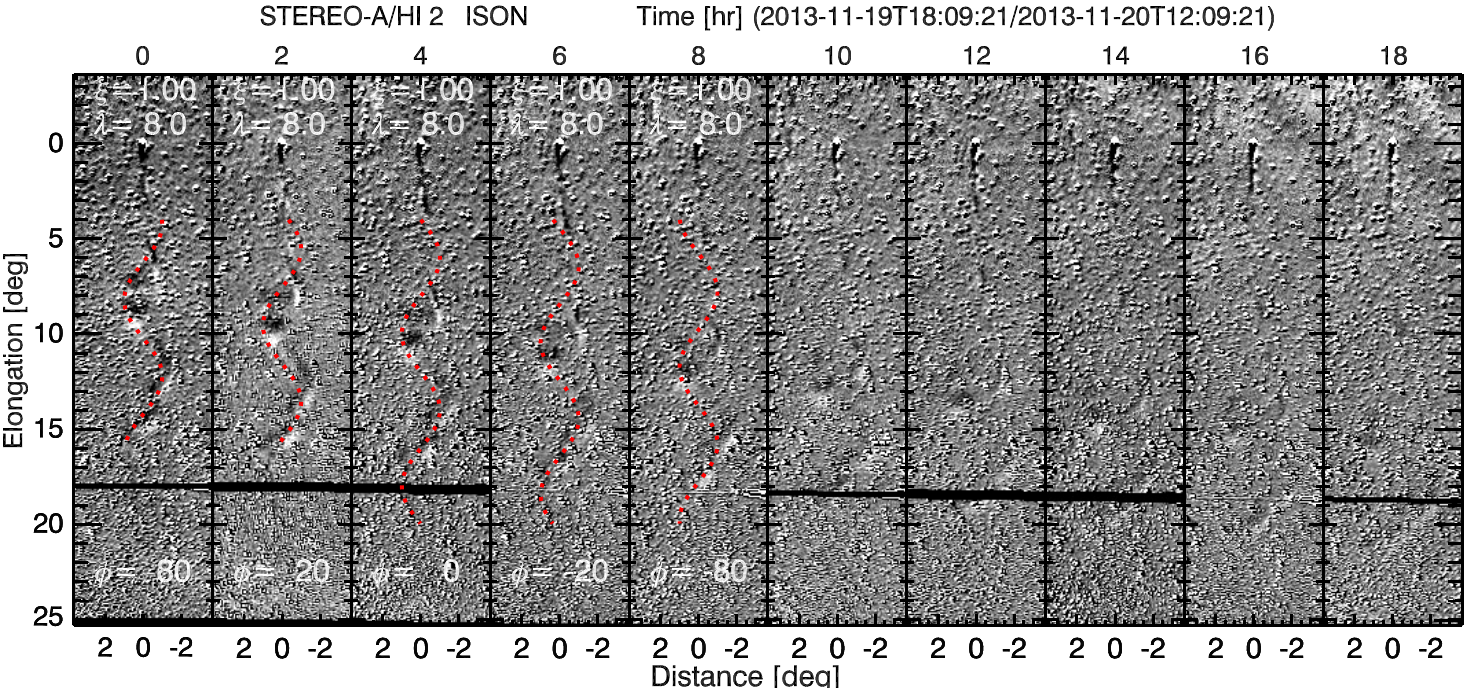}\\
		\includegraphics[width = 0.5\textwidth]{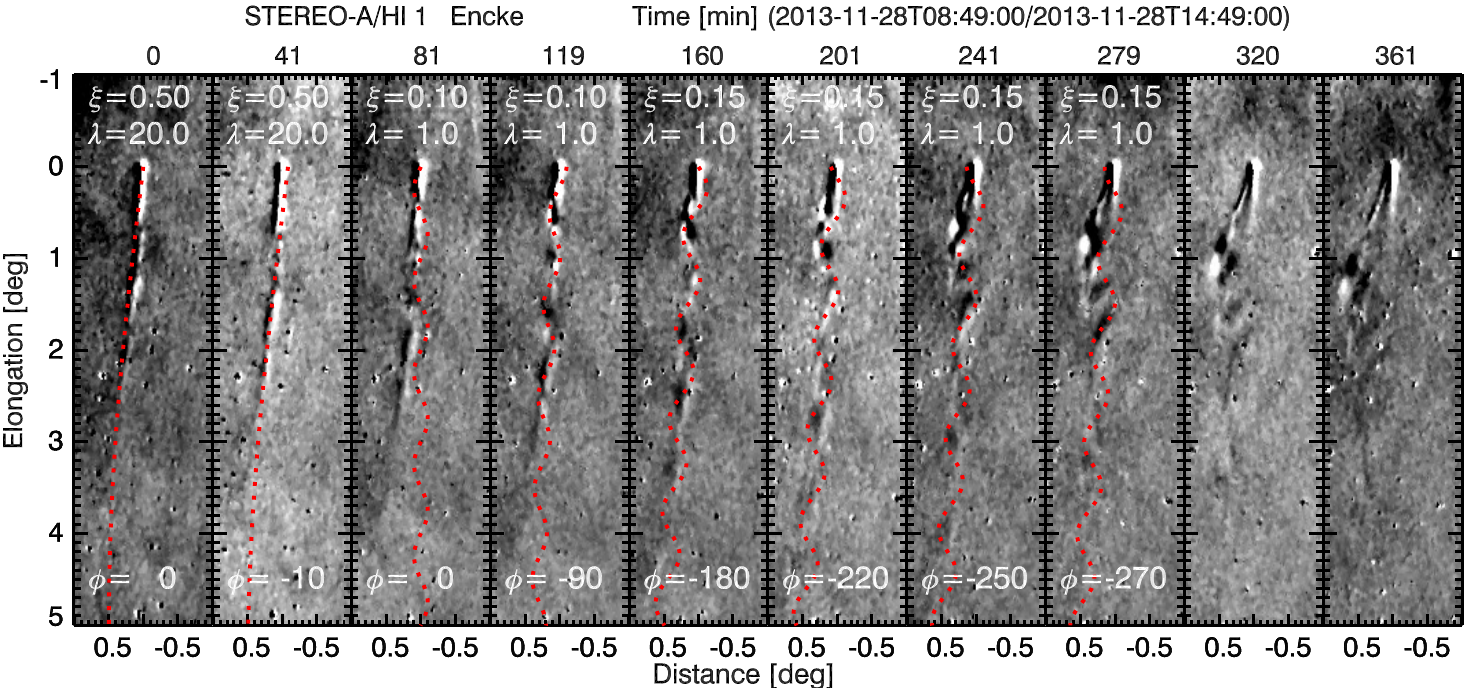} &
		\includegraphics[width = 0.5\textwidth]{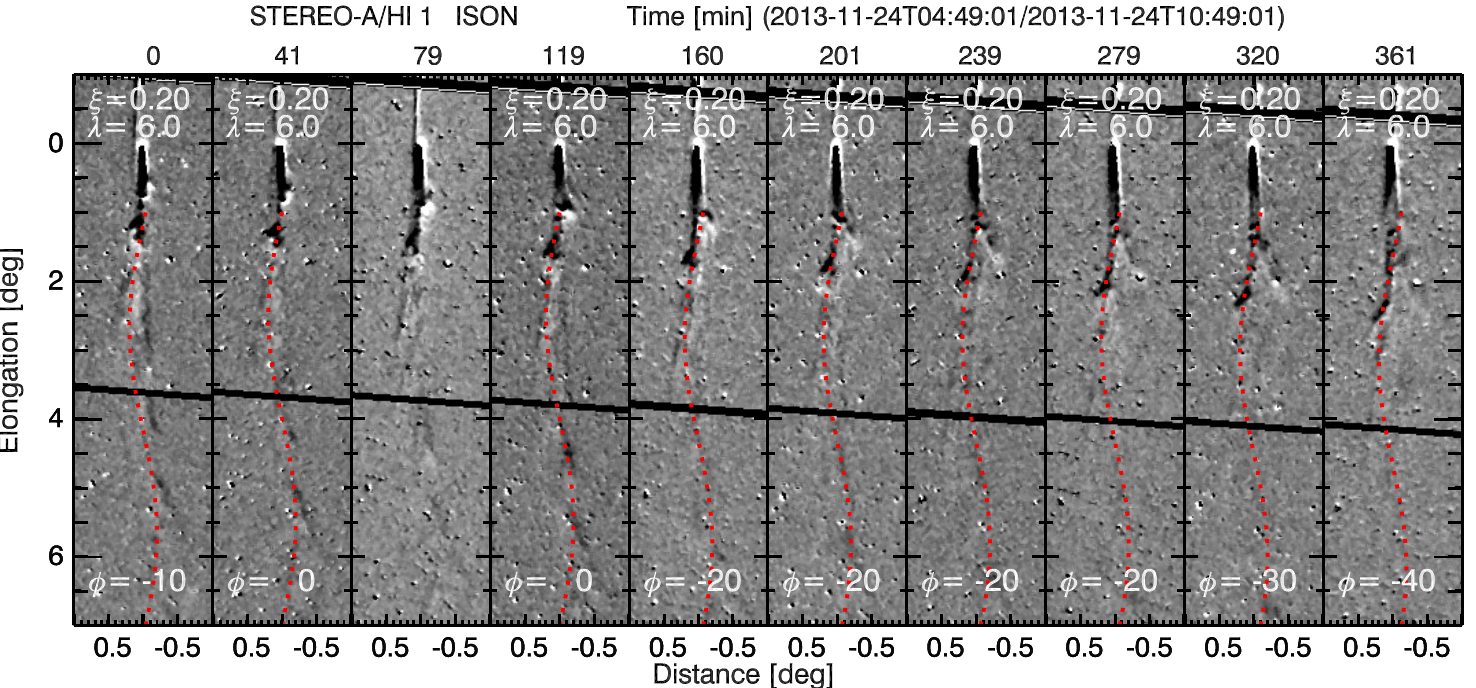} \\
			\end{tabular}
	\caption{Top: Time distance maps for Encke and ISON showing the manually determined data points tracking the oscillations. Middle and bottom: Sequences of 10 consecutive frames for the comet Encke (left) and ISON (right) with some sinusoids marking propagating waves along the tail. 
	Values of amplitude, wavelength, and phase of the oscillations are reported on each single frame. Numbers on the top horizontal axis for each frame shows the time-steps in min from the first frame of the sequence. }
	\label{fig_period}
\end{figure*}

	Periods for the oscillations of the tail are determined by TD maps constructed with a vertical slit located 
	at a given distance (e.g., 40, 50, 60, ... pixels) from the comet's coma in the processed running difference sub-images.
	An example is given in Fig. \ref{fig_period}, where the TD
	maps for the comet Encke and ISON in HI-1A are extracted from a slit located at 50 pixel (1.0 deg
	$\approx 2.5\times10^6$ km) from the comet's coma. The manually-determined points are fitted with a
	sinusoidal function plus a linear function to take into account any possible deviation from the local 
	zero:
	\begin{equation}
		y = y_0 + \xi \sin \left( \frac{2\pi}{P} t + \phi\right) + C t.
	\end{equation}
	
	In order to detected different possible regime of oscillations, we divided the obtained times series in
	consecutive intervals of 10, 20, 30, 60, and 90 hours.
	Periods ranging between 5 and 20 hr have been measured both for Encke and ISON.
	We only considered good estimates those periods having a relative error $\sigma_P/P < 30\%$ obtained from
	fittings with $\chi^2< 10$. The amplitude of the fitted oscillations are also scaled by the factor $d_\mathrm{C}/
	d_\mathrm{Sun}$ in order to account for the distance comet-observer.

\subsection{Estimation of the Strouhal numbers}
\begin{figure*}[htpb]
\centering
	\begin{tabular}{c c}
		\includegraphics[width = 8 cm]{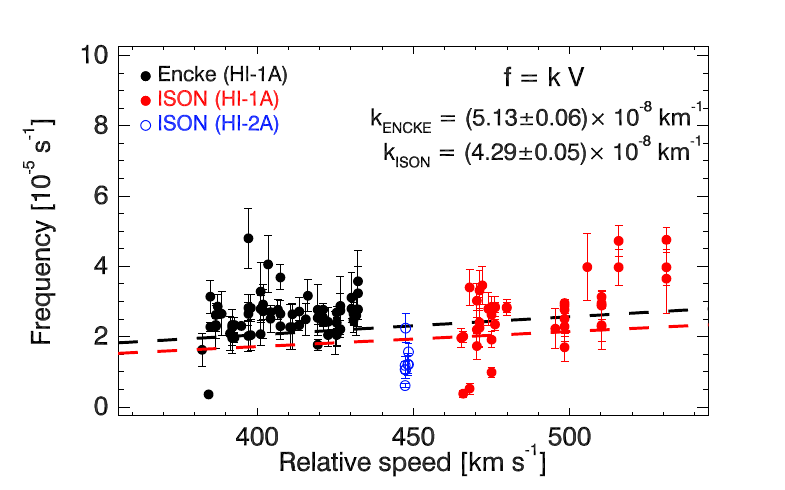} &
		\includegraphics[width = 8 cm]{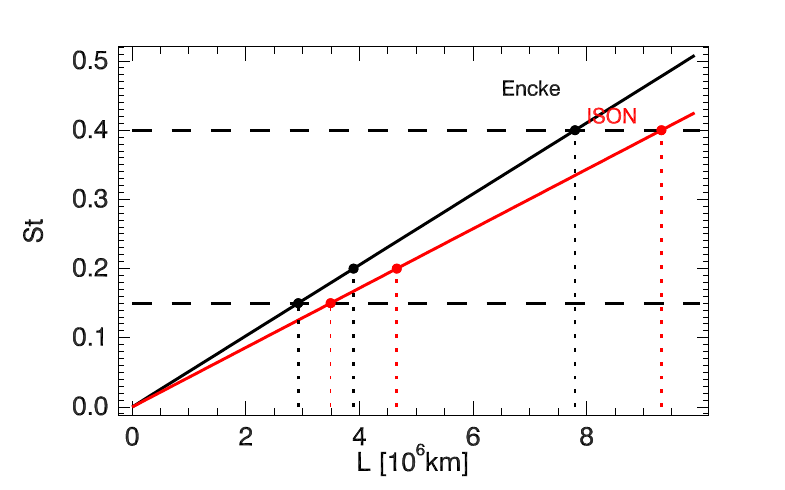} \\
		\includegraphics[width = 8 cm]{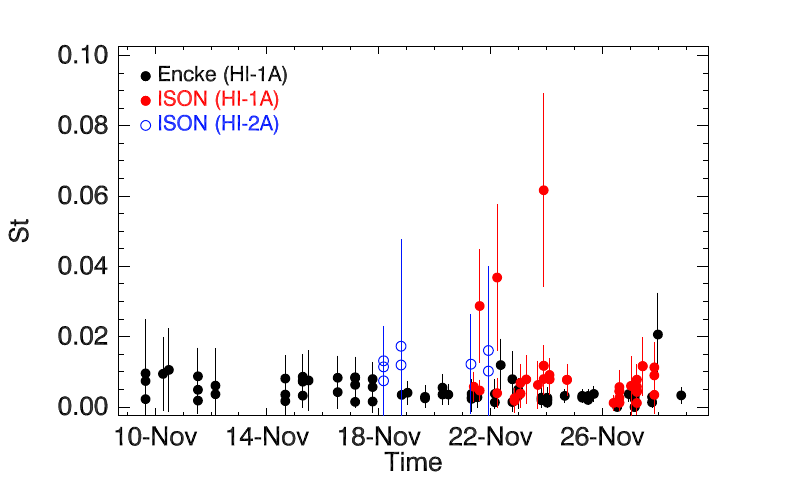} &
		\includegraphics[width = 8 cm]{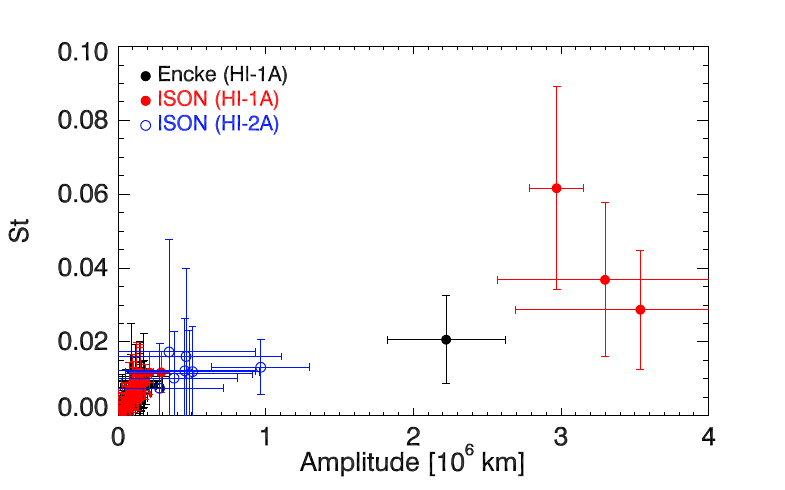} \\
		\includegraphics[width = 8 cm]{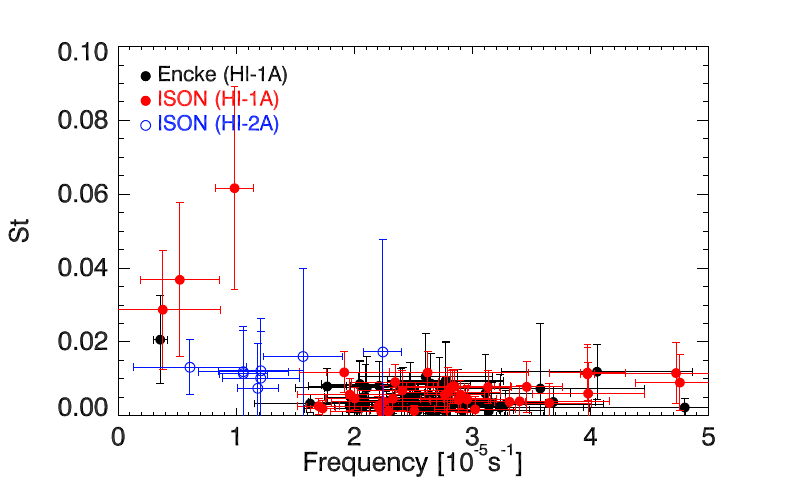} &
		\includegraphics[width = 8 cm]{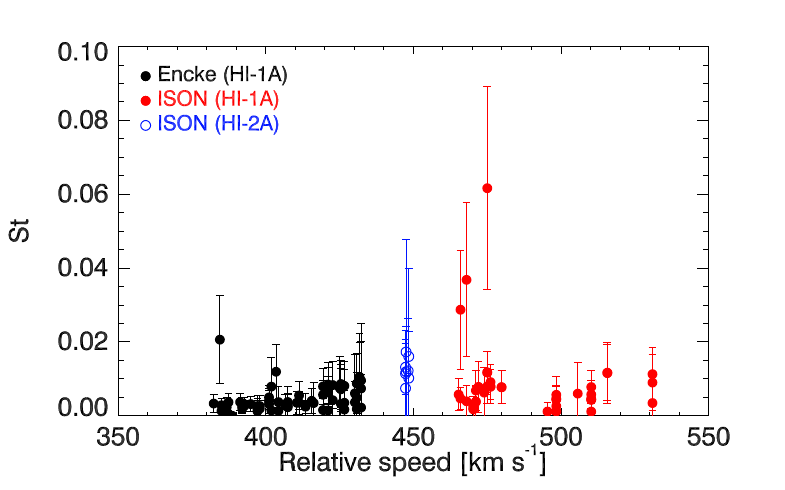} \\
	\end{tabular}
	\caption{Top-left: scatter plot of frequency vs. relative speed (right) for Encke in HI-1 (black dots), and ISON in HI-1 (red) and HI-2 (blue). The slope of the fitting line is related to the Strouhal number. Top-right: extrapolation of the Strouhal number with respect to the coma size $L$. Middle: plots showing the Strouhal number of each data point vs. time (left), and vs. the oscillation amplitude. Bottom: plots of $St$ vs. the frequency $f$ (left), and vs. the relative speed $V$.}
	\label{fig_st}
\end{figure*}

\begin{figure*}[htpb]
\centering
	\begin{tabular}{c c c}
\includegraphics[width=0.33\textwidth]{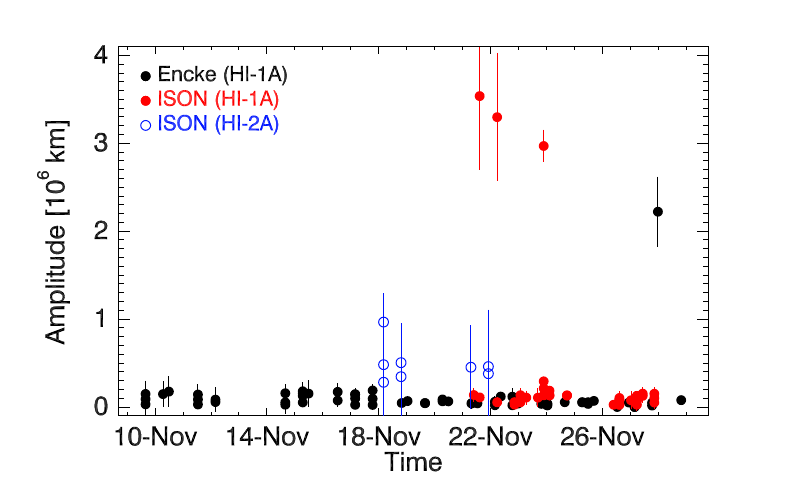} &
\includegraphics[width=0.33\textwidth]{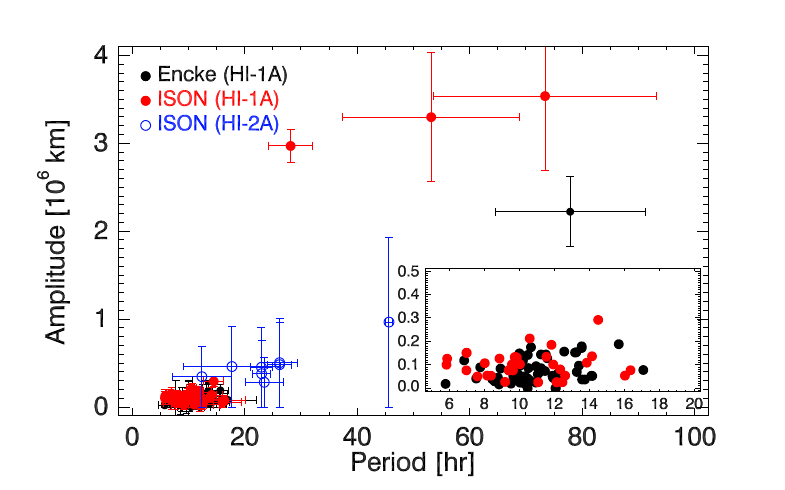} &
\includegraphics[width=0.33\textwidth]{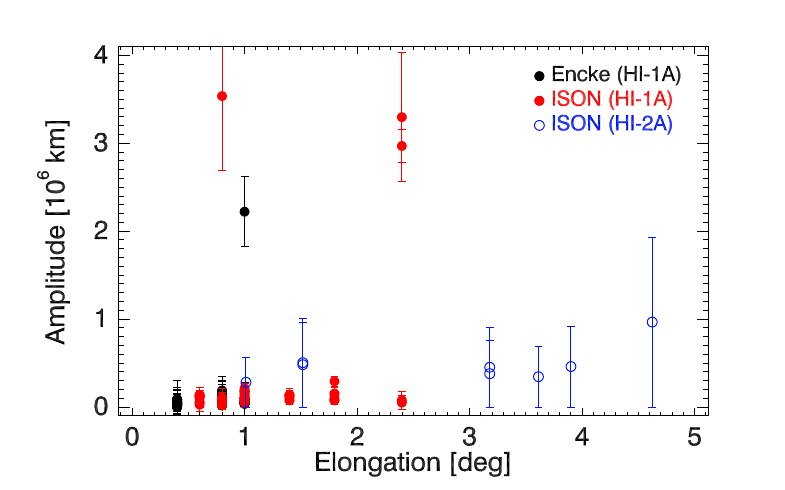} \\
	\end{tabular}
	
	\caption{Variation of the oscillation amplitude vs. time,  vs. period of the oscillations (the inset plot magnifies the inner region containing Encke and ISON's data points in HI-1A), and vs. distance along the cometary tail. In black Encke's data points, in red and blue ISON's data points from HI-1 and HI-2, respectively.}
	\label{fig_ampl}

\end{figure*}

By relating the estimated frequencies $f = 1/P$ and the corresponding relative speeds $V$,
 we fitted the data points with a linear function $f = k V$, where $k =  St/L$ (Fig. \ref{fig_st}-top left). When doing this, we have not considered a time lag between the value of $V$ and $f$, since some delay is reasonably expected between the times when the halo encounters a given speed, the vortex is formed, advected with a given phase speed and then measured at a given distance from the halo. Hence, the data points should be moved towards lower values of speed, but this would be a minor correction. Given $k$ and $L$, we find Strouhal numbers of the order of $10^{-3}$, which are 
considerably small, with some values between 0.02--0.1 (Fig. \ref{fig_st}).
In hydrodynamics $St\approx0.2$ for a very broad range of parameters, which should be
associated with $f^{-1}\approx0.3$ hr for $L=10^5$, and $V=400$ km s$^{-1}$. By extrapolating $St$ at higher values 
of L using the determined coefficients $k$, we find that values of $St\approx0.15-0.4$ are obtained for
 $L\approx2.5-7.2 \times 10^6$ km (Fig. \ref{fig_st}-top right), which are very large, even if in agreement with
 typical scale lengths. For example, Ulysses crossed the tail of the comet Hyakutake in 1996 at a distance of 
3.8 AU from its nucleus, and measured a diameter of $\sim 7 \times 10^6$ km \citep{Jones2000}. However, such a 
value is improbable in the proximity of a nucleus (indeed the tail undergo cross-sectional expansion), and an upper limit value can 
be reasonably considered as $1\times10^6$ km (assuming that the outermost layers are indeed not detected with 
HI-1), which should represent the size of the overall draped magnetic structure around the cometary nucleus. 
On the other hand, a halo  of hydrogen is developed around comets with a diameter even larger than
 the Sun.
The parameter $L$ can be associated with the size of the shed vortices, which undergo expansion due to diffusivity. Our oscillations are measured at prescribed distances from the coma, where the oscillation amplitudes have in some few cases values of the order of 10$^6$ km, which, however, markedly deviated from the sample distribution (Fig. \ref{fig_ampl}).
 
In the middle and bottom panels of Fig. \ref{fig_st} we plot the values of $St$ for each data point (we used the local oscillation amplitude as $L$), 
showing how it changes with time, the local oscillation amplitude, the frequency $f$, and the relative speed $V$. 
Some extreme values are larger than 0.02, corresponding to the extreme amplitude values mentioned previously,  but in general the cloud of points lies under $St\approx0.01$.

\section{Discussion and conclusions}
\label{sec:disc}

The small values of the estimated Strouhal number raise the questions of whether the observed kink-like oscillations of the plasma tail are induced by the solar wind variability (e.g. due to CMEs), or associated with vortex shedding like phenomena. 

In the former case, the observed oscillations would be not natural, and it would require the oscillation in the wind to be, 
as we observed in the tail, monochromatic and of a large amplitude. In addition, for the excitation of the
oscillations of the kink symmetry in the tail, the oscillation in the wind should be of the same kink symmetry 
and we are not aware of this. Thus, we should disregard this interpretation on this basis. Another option 
is that the oscillations in the solar wind excite natural modes of the tail by resonance. 
In this case the oscillations in the wind could be broadband, and only the resonating harmonics take part in the excitation of the natural modes in the tail (i.e. a harmonic oscillator driven by a broadband force). However, in this scenario the tail oscillation should grow gradually, and also, variations of the phase of the induced oscillation in the tail would be expected, which we do not see either.

\begin{figure}[htpb]
\centering
	\includegraphics[width = 0.5\textwidth]{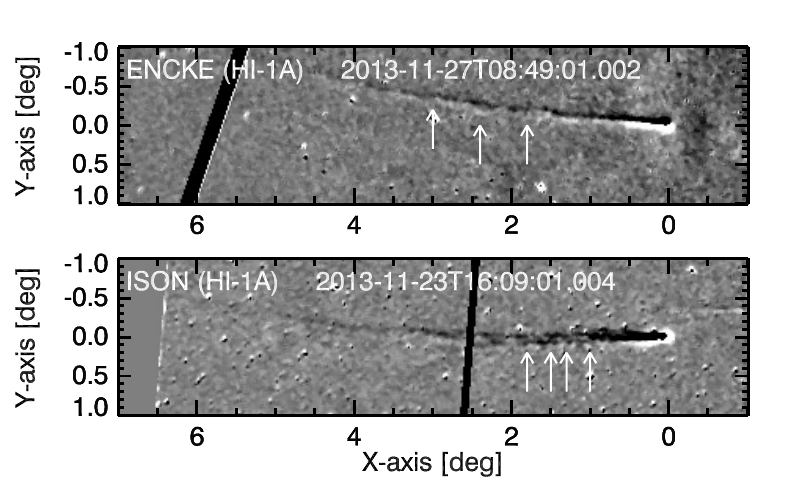}
	\caption{Snapshots of Encke and ISON from HI-1A with some small-scale perturbations of the tails highlighted by the white arrows.}
	\label{encke_ison_single}
\end{figure}

\begin{figure*}[htpb]
\centering
	\begin{tabular}{c c}
		\includegraphics[width=.45\textwidth]{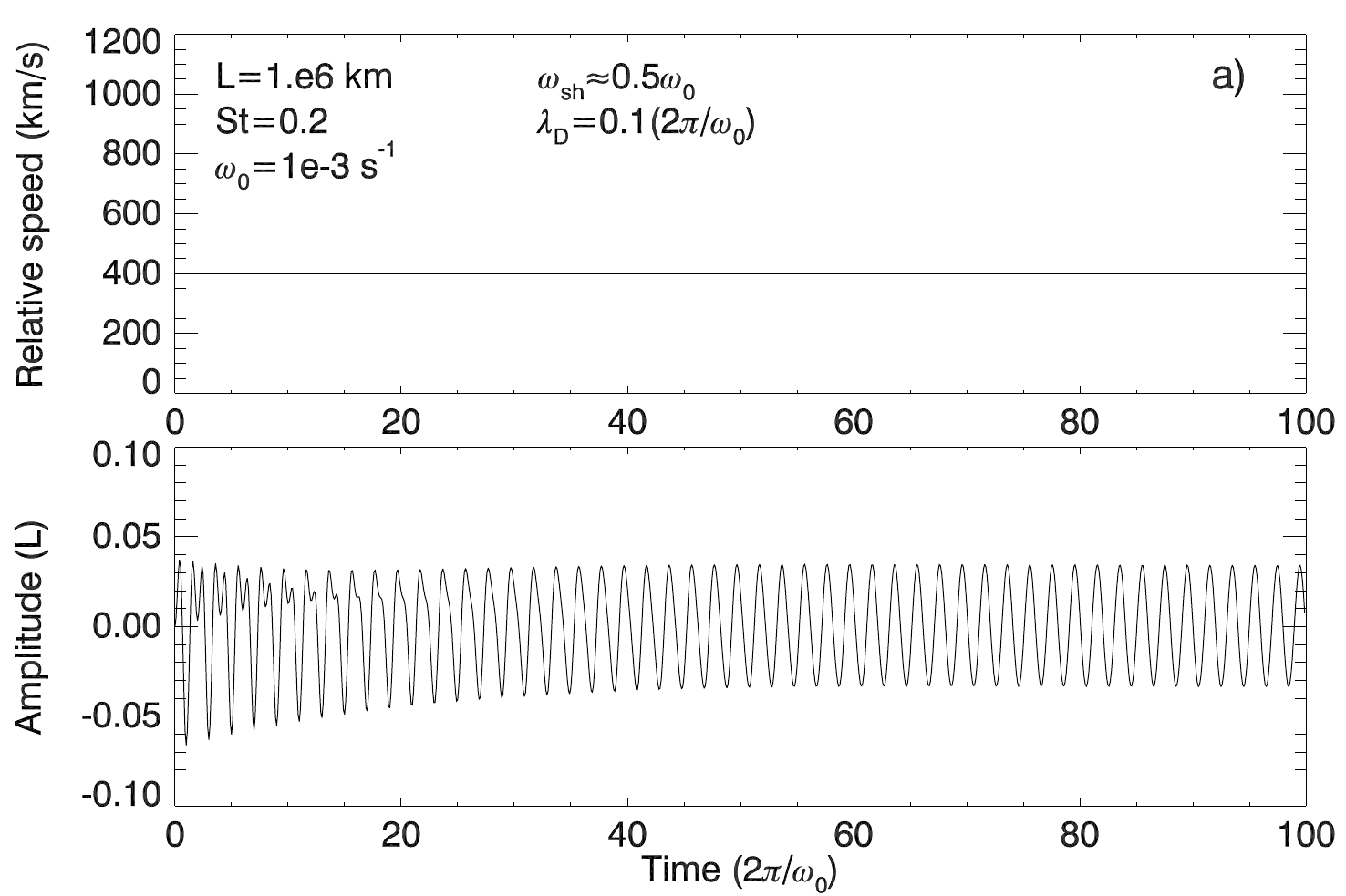} &
		\includegraphics[width=.45\textwidth]{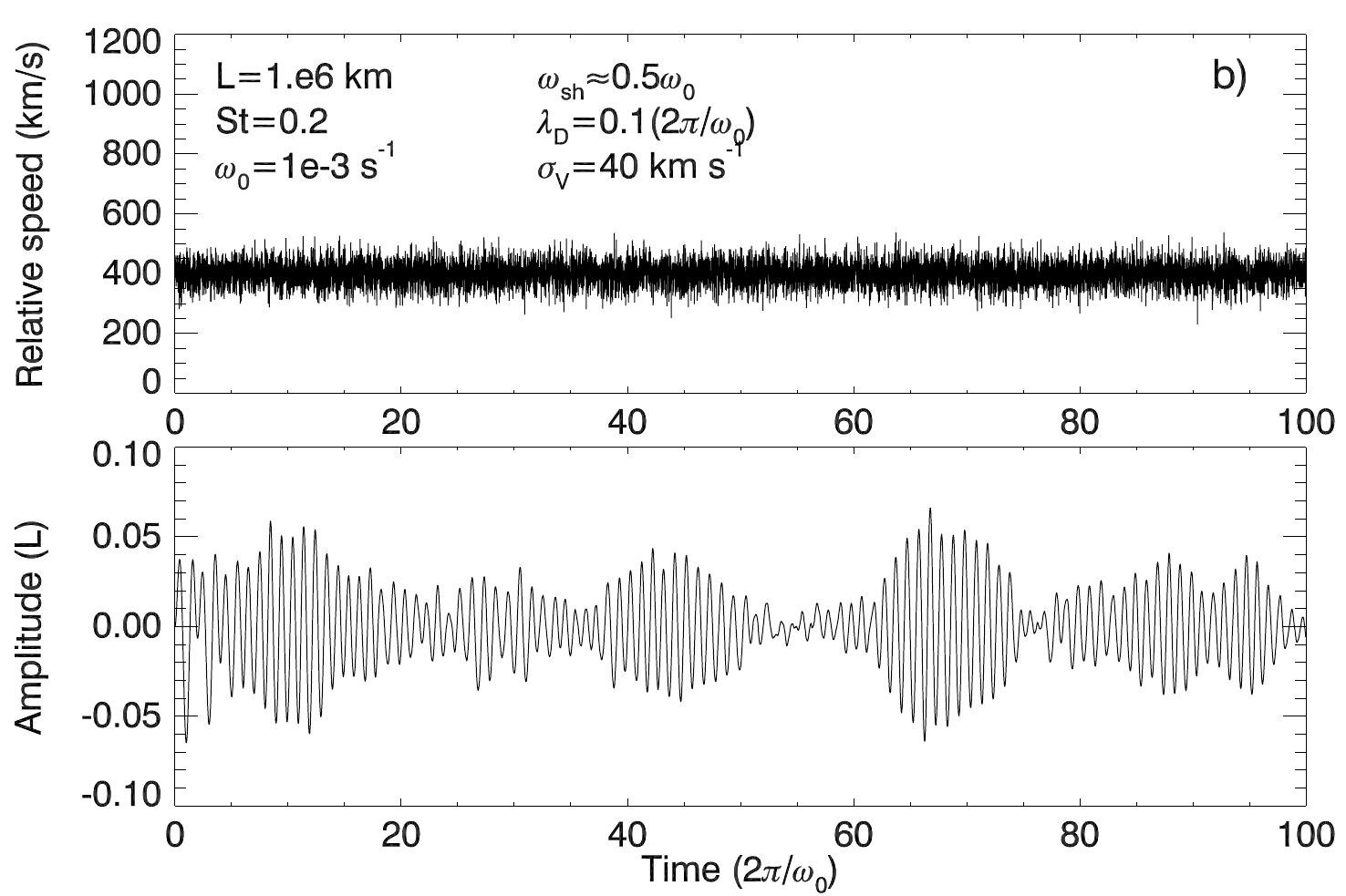}  \\
		\includegraphics[width=.45\textwidth]{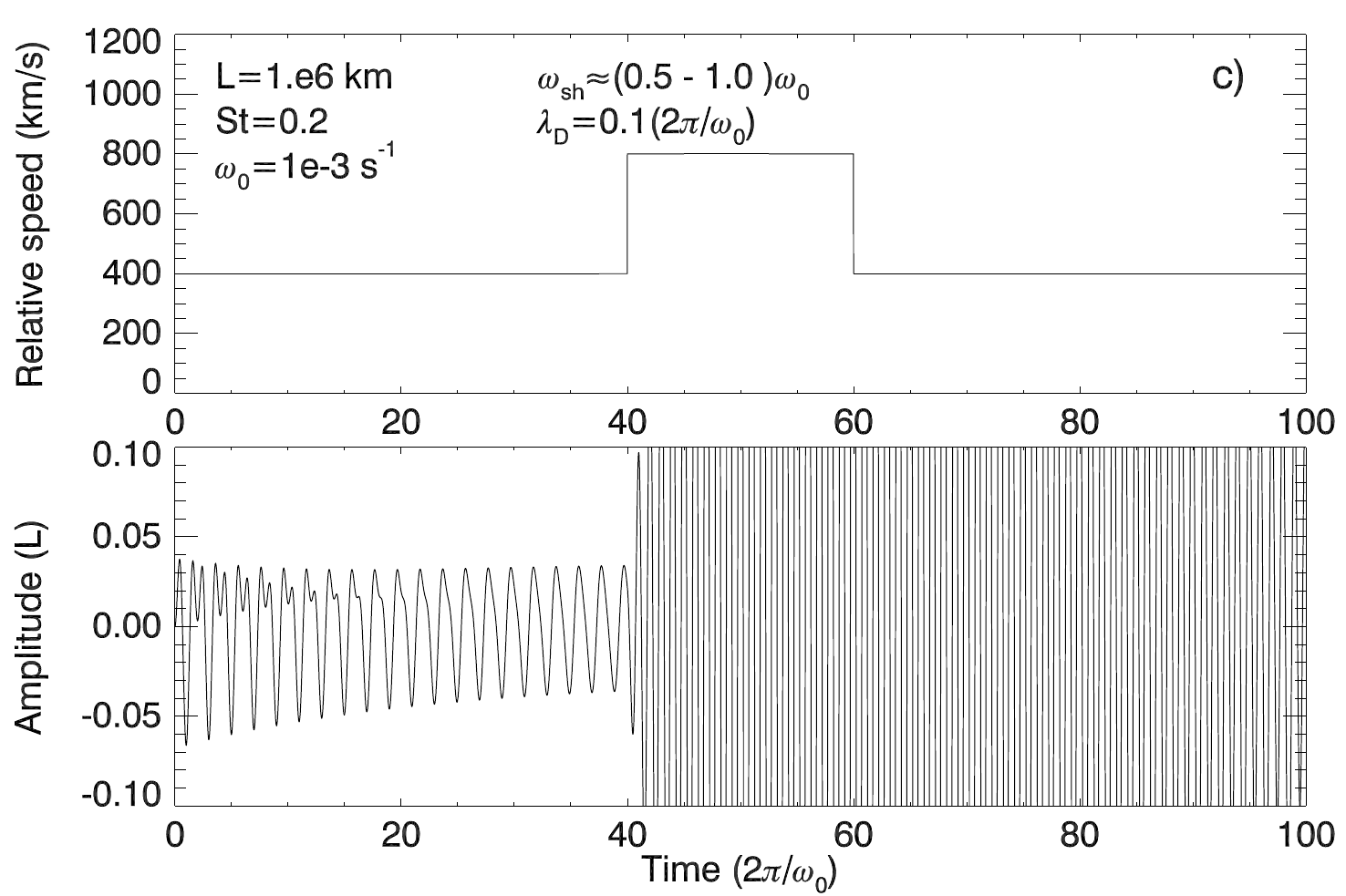} &
		\includegraphics[width=.45\textwidth]{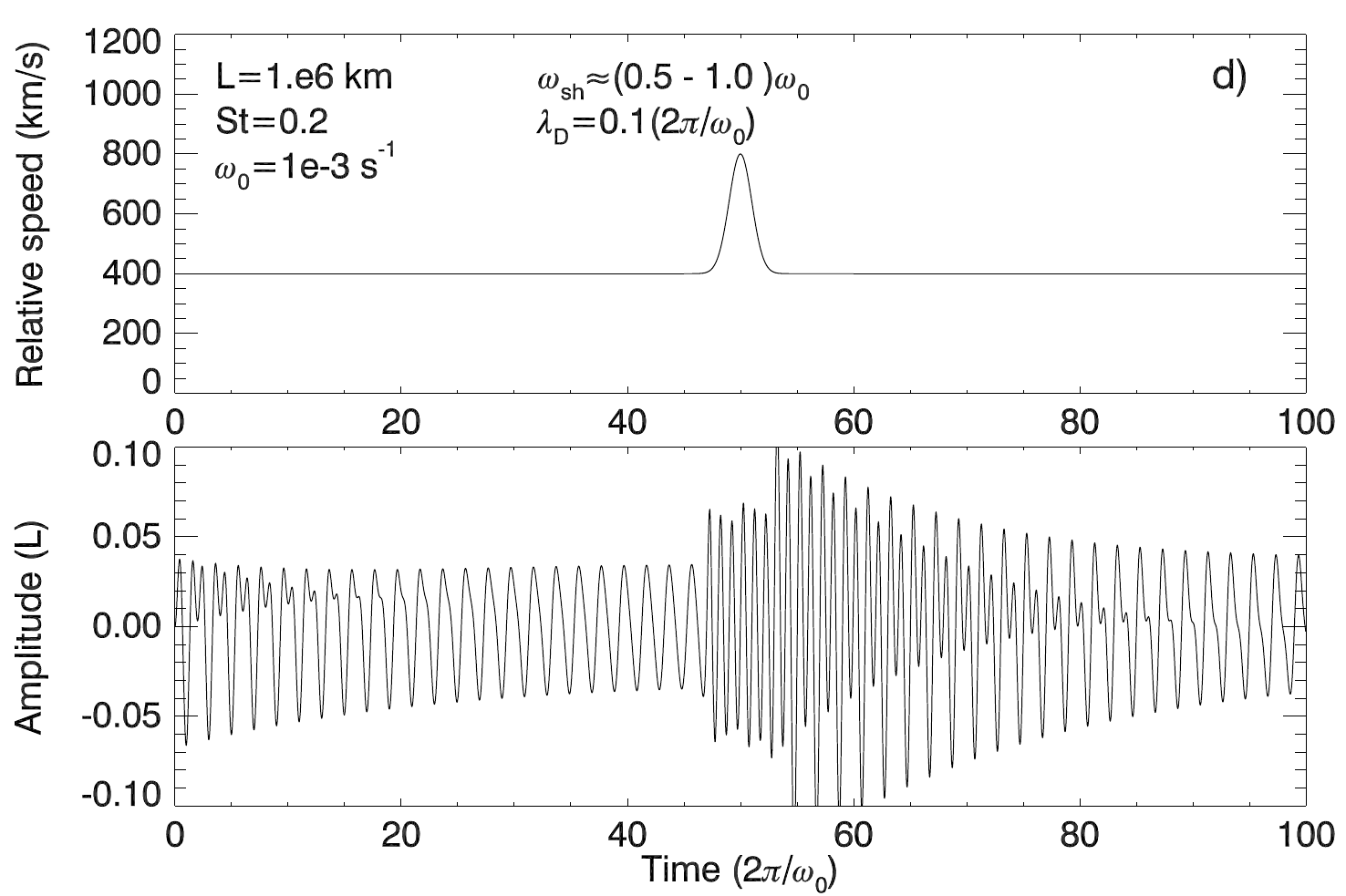} \\
		\includegraphics[width=.45\textwidth]{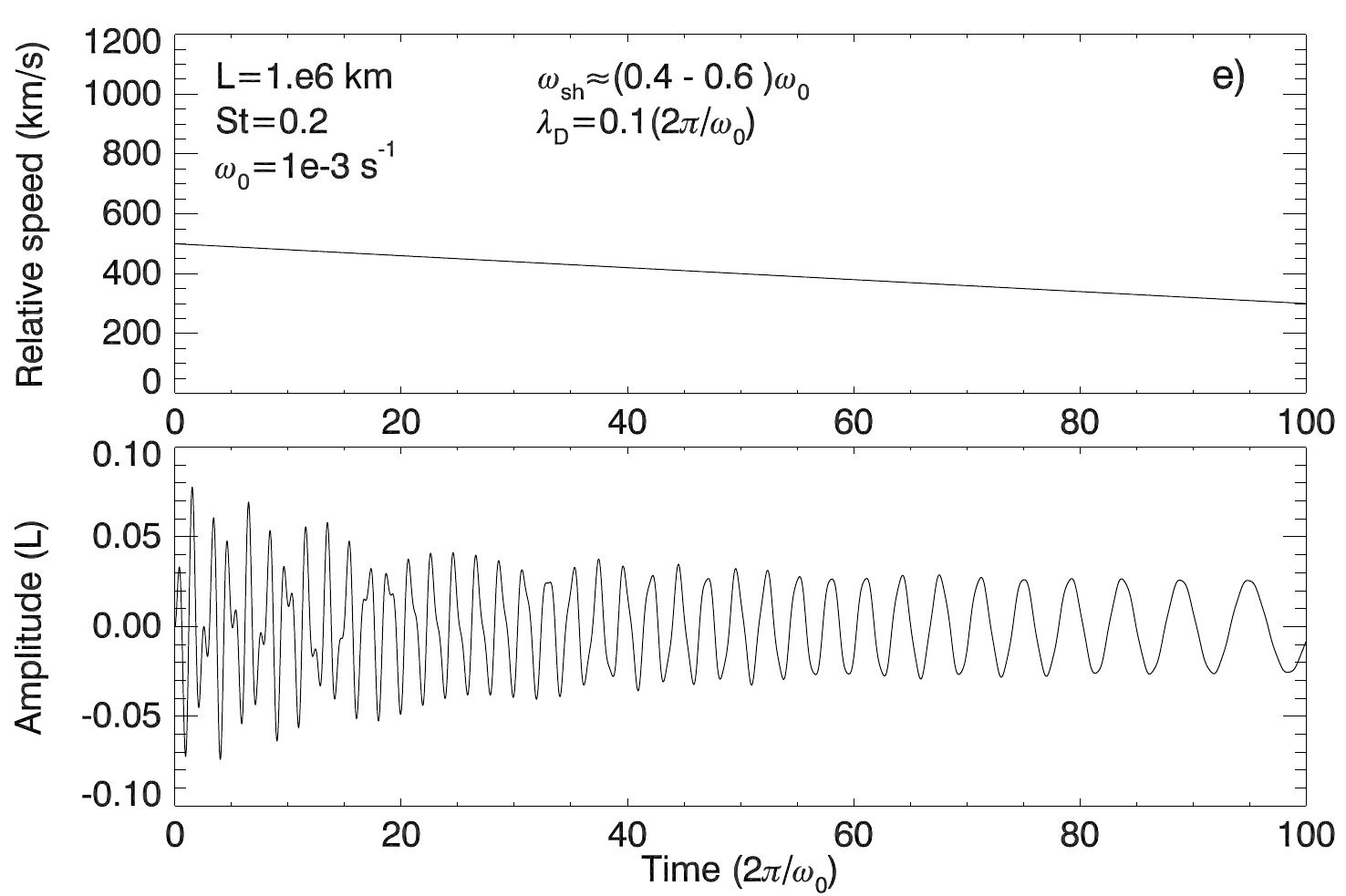} &
		\includegraphics[width=.45\textwidth]{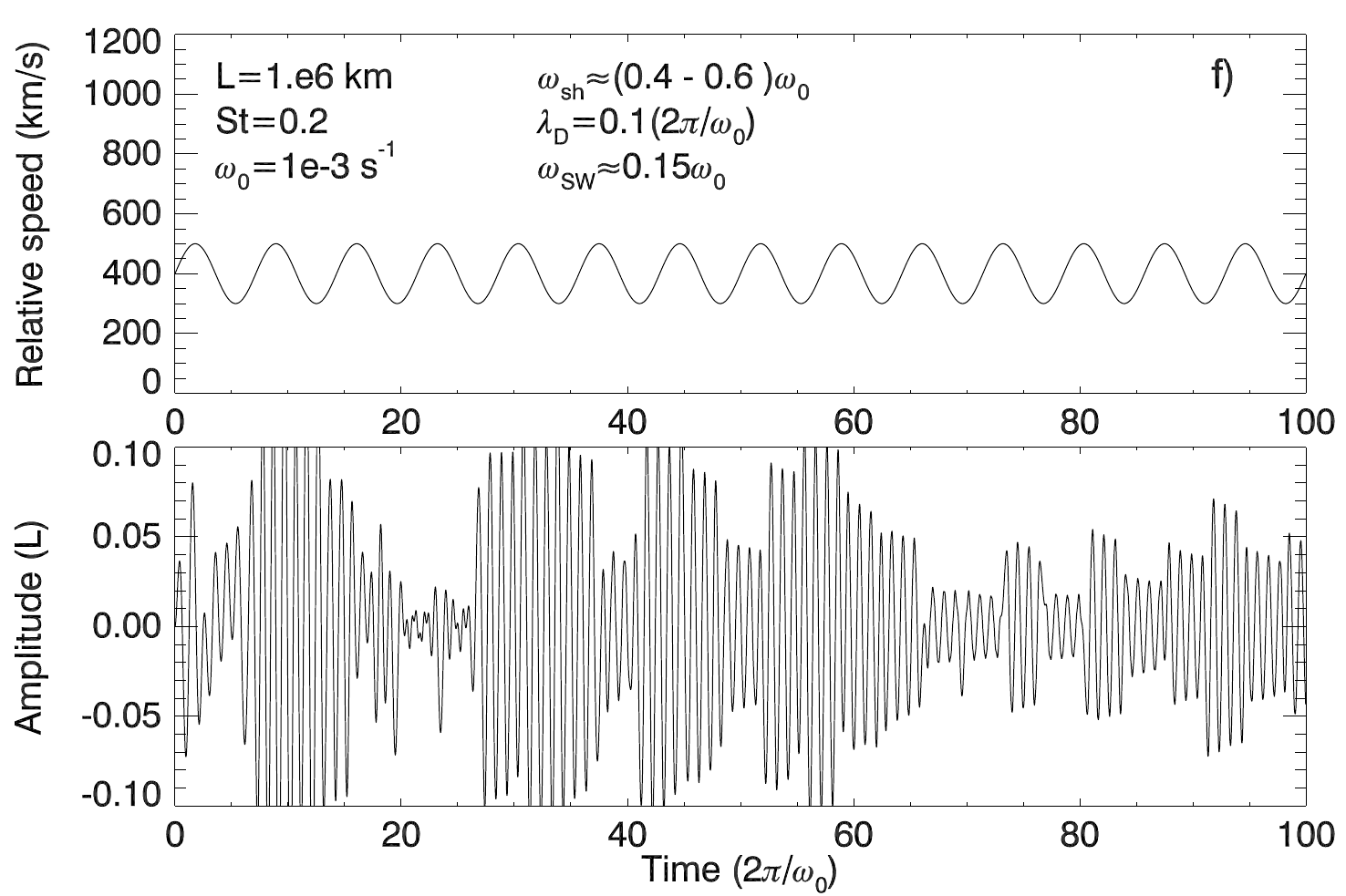} \\
	
	\end{tabular}
	\caption{Numerical solutions of the differential equation \eqref{eq:harm_oscill} for different types of ideal solar wind profiles.}
	\label{fig:model}	
\end{figure*}

In the latter case, the tail oscillations may be associated with a breakdown of the proper K\'arm\'an vortex street into a secondary structure 
\citep{Johnson2004,Dynnikova2016}. Something similar also appears in the simulations of \citet{Gruszecki2010} 
(see their Fig. 1), where the entire vortex street has an oscillatory structure, with a wavelength 
4-5 times larger than the vortex size (presumably the period should be 4-5 times longer than the vortex 
shedding period, if we assume identical phase speed in both regimes). In such a case, our estimates for $St$ should be corrected by the same factors, hence $St$ values will range between $\approx0.02-0.3$. On the other hand, small-scale perturbations appear in the tails of Encke and ISON (white arrows in Fig. \ref{encke_ison_single}), which however have not properly considered in the present study because of limitations in the spatial resolution of the instruments and intensity contrast.
It is interesting to notice that the comet tail-solar wind flow can be modelled in terms of a damped driven harmonic oscillator

\begin{equation}
	y'' + \lambda_\mathrm{D} y' + \omega_0^2 y = F(t) \cos(\omega_\mathrm{sh}(t) t),
	\label{eq:harm_oscill}
\end{equation}
with $y$ the displacement, $\omega_0$ the natural frequency of the magneto-acoustic mode of the tail (which can be assimilated to a plasma cylinder), $F(t)$ 
the amplitude of the external force, $\omega_\mathrm{sh}(t) = 2 \pi St V(t)/(nL)$ the vortex 
shedding pulsation (we added an artificial factor $n$ to model the possible 
contribution from a low-frequency mode for $n>1$, $n=1$ would simply 
correspond to a pure vortex shedding mode), and $\lambda_\mathrm{D}$ the damping factor. 
The characteristic of the natural magneto-acoustic frequency $\omega_0$, in the presence of steady flows internally or externally to the magnetic tube (in our case, the solar wind would play the role of the external flow) can be modified as shown in \citet{Nakariakov1995}, and also be suppressed under particular conditions.
In the non-resonant regime ($\omega_\mathrm{sh}\ne\omega_0$), the period of the oscillator is prescribed 
by the external driving frequency, which however has a variable nature because of the dependance on the 
solar wind speed by $V$. Sudden increases of $V$, e.g. due to the passage of a coronal mass ejection, 
may lead to an abrupt change in the frequency regime or to a disconnection tail event if resonance is 
achieved ($\omega_\mathrm{sh} \approx \omega_0$). Some examples are shown in Fig. \ref{fig:model} with the tail displacement solution from \eqref{eq:harm_oscill}, given values of $L$, $St$, $\lambda_\mathrm{D}$, $\omega_0$, and for constant value of $F(t)$ and $n = 1$. We used some test-functions for the relative speed profile (e.g. constant profile at 400 km s$^{-1}$ (a), with added Gaussian noise (b), square function (c), with a Gaussian peak (d), linear trend (e), and sinusoidal profile (f)). When $V$ reaches 800 km s$^{-1}$, the vortex shedding pulsation $\omega_\mathrm{sh}$ equalises the natural pulsation $\omega_0$, and the amplitude of the oscillations increases because of resonance (cases c, d). In other cases (b,f), we observe the formations of beats with frequency of occurrence much lower than the natural frequency of the oscillator.     
In addition, damping effects have an important role in shaping the oscillations. 
Formation of vortices in plasma are strongly affected by magnetic fields. Numerical simulations 
of vortex shedding in a plasma flow past a cylinder have been undertaken by \citet{Gruszecki2010} with a 
magnetic field strictly perpendicular to the plane of the flow (or in other words with the magnetic field direction coincident with the direction of the vorticity vector). In this case, the induced oscillation period is determined by the Strouhal number similar to the pure hydrodynamic case. On the other hand, it is well know that a parallel magnetic field has a stabilising effect on unstable modes (e.g. the KH instability) because of the appearance 
of a Lorentz force \citep[pp. 45-51 of][]{Biskamp}.    
With the presence of a component for magnetic field parallel to the tail, the condition for stability in ideal MHD is determined by the local Alfv\'en speed $V_\mathrm{A}$ and the jump in velocity $\delta V$ across a sheet (i.e., the cometary plasma tail in our case; ideally $\delta V$ would be equal to the unperturbed $V_{SW}$, since one can assume velocity 0 in the middle of the plasma tail), that is $V_A > \frac{1}{2} |\delta V|$ \citep[p. 50 of][]{Biskamp}. The condition for the stability discussed above can be exploited for the determination of the local magnetic field.   

From Fig. \ref{fig_period} we note that the tail structure is often straight up to few degrees of elongation from the coma, maybe because the local $\delta V$ does not exceed the local Alfv\'en speed. 
This effect on the structuring of the tail 
is very evident for the comet ISON when it is moving out of the FoV of HI-1A towards the perihelion: the comet is getting even closer to the Sun, moving in a region where the local magnetic field is increasing, and 
consequently the local $V_A$ is growing. On the other hand, the increase of $V_A$ should be attenuated 
by the increase in the plasma density. 

Direct measurements in cometary tails with spacecraft show that the interplanetary field is draped around 
the comet nucleus, with magnetic field lines of opposite polarities at the side of a neutral sheet (which 
correspond to the plasma tail) \citep[see Figs. 8.22 and 8.23 in][]{Kivelson}, as observed in the case of the comet Giacobini-Zinner \citep{Malara1989a} and Hyakutake \citep{Jones2000}. Such a configuration, in 
addition to the KH instability, is also inclined to tearing mode instability, which may break the downstream 
magnetic structure of the cometary tail in a series of islands along the neutral plane, which can be observed 
in the form of small-scale plasma condensations \citep{Malara1989b}, or being responsible of a tail disconnection 
event \citep{Vourlidas2007}. However, these perturbations are of the sausage symmetry, and hence are different from the kink oscillations detected in this study. Major details on all these aspects can only come by targeted MHD simulations.

Altough we have not provided conclusive evidences, we suggest that oscillations in cometary tails may be explained in terms 
of the interaction between the comet and surrounding environment by vortex shedding phenomena. Furthermore, the presence of eddies has been recently shown in a study of the comet Encke 
during its perihelion in 2007 \citep{DeForest2015}, with an energy content enough to heat the 
solar wind plasma. Certainly, there are big differences in the nature and composition of the tails of Encke 
and ISON, that we investigated in this study. While Encke is a very stable comet, ISON experienced several 
explosive fragmentation \citep{Sekanina2014,Keane2016}, which may have perturbed that tail. The lack of a 
coherent nucleus \citep{Knight2014}(hence a fully developed coma and magnetic cavity) may explain the lack of 
oscillations after the ISON's perihelion. Using comets as probes of the inner heliosphere is additionally promising for 
inferring local plasma properties. For example, values of $St=0.15-0.2$ for $L> 2 \times 10^6$ km in a pure 
hydrodynamic flow would be associated with $Re \approx$ 300-400 in the case of a sphere \citep{Sakamoto1990}, 
which in turn would correspond to an effective kinematic viscosity of the order of 
$\nu = 10^6$ km$^2$ s$^{-1}$ . Estimates of the kinematic viscosity of $10^9$ km$^2$ s$^{-1}$ 
(also called large-scale eddy viscosity for the solar wind) are given in \citep{Verma1996} based upon 
theoretical assumptions. The discrepancy could be also attributed to the increase in the effective viscosity caused by the plasma micro-turbulence. 
 


\begin{acknowledgements}
STEREO-HI data are courtesy of the UK Solar System Data Centre. 
VV and GN acknowledge support from the URSS scheme at the University of Warwick. GN and VB acknowledge support of the CGAUSS (Coronagraphic German And US Solar Probe Plus Survey) project for Parker Solar Probe/WISPR by the German Space Agency DLR under grant 50 OL  1601. VMN acknowledges the funding by STFC consolidated grant ST/P000320/1. KB was supported by the NASA-funded Sungrazer Project. GN would also thank the members of CFSA at the University of Warwick and the Astrophysics group at the University of Calabria for useful comments and discussion after the presentation of this subject.
\end{acknowledgements}

\bibliographystyle{aa}
\bibliography{references} 


\end{document}